\documentclass[twocolumn,a4paper,showpacs,preprintnumbers,amsmath,amssymb,pre]{revtex4-1}
\usepackage{graphicx}% Include figure files
\usepackage{dcolumn}% Align table columns on decimal point
\usepackage{subfigure}
\usepackage{bm}% bold math
\usepackage{setspace}
\usepackage{fancyhdr}
\usepackage{amsmath}
\usepackage{color}
\usepackage{epstopdf}
\usepackage{psfrag}

\begin{document}

\title{Spatial distribution of heat flux and fluctuations \\ in turbulent Rayleigh-B\'{e}nard convection}

\author{Rajaram Lakkaraju$^{1}$}
\author{Richard. J. A. M. Stevens$^1$}
\author{Roberto Verzicco$^{1,2}$}
\author{Siegfried Grossmann$^3$}
\author{Andrea Prosperetti$^{1,4}$}
\author{Chao Sun$^1$}
\author{Detlef Lohse$^1$}

\affiliation{$^1$Faculty of Science and Technology, Mesa+ Institute and J. M. Burgers Center for Fluid Dynamics, University of Twente, 7500AE Enschede, The Netherlands}
\affiliation{$^2$Department of Mechanical Engineering, University of Rome `Tor Vergata', Via del Politecnico 1, 00133 Rome, Italy}

\affiliation{$^3$Department of Physics, University of Marburg, Renthof 6, D-35032 Marburg, Germany}

\affiliation{$^4$Department of Mechanical Engineering, Johns Hopkins University, Baltimore, MD 21218, USA}

\date{\today}

%%%%%%%%%%%%%%%%%%%%%%%%%%%%%%%%%%%%%%%%%%%%%%%%%%%%%%%%%%%%%%%%%%%%%%%%%%%%%%%%%%%%%%%%%%%%%%%%%%%%

\begin{abstract}

We numerically investigate the radial dependence of the velocity and temperature fluctuations and of the time-averaged heat flux $\overline{j}(r)$ in a cylindrical Rayleigh-B\'{e}nard cell with aspect ratio $\Gamma=1$ for Rayleigh numbers $Ra$ between $2 \times 10^6$ and 
$2\times 10^{9}$ at a fixed Prandtl number $Pr = 5.2$. The numerical results reveal that the heat flux close to the side wall is larger than in the center and that, just as the global heat transport, it has an effective power law dependence on the 
Rayleigh number, $\overline{j}(r)\propto Ra^{\gamma_j(r)}$. The scaling exponent $\gamma_j(r)$ decreases monotonically 
from $0.43$ near the axis ($r \approx 0$) to $0.29$ close to the side walls ($r \approx D/2$). The effective exponents near 
the axis and the side wall agree well with the measurements of Shang et al. (Phys.\ Rev.\  Lett.\ \textbf{100}, 244503, 2008) and the predictions of Grossmann and Lohse (Phys.\ Fluids \textbf{16}, 1070, 2004). Extrapolating our results to large Rayleigh number would imply a crossover at $Ra\approx 10^{15}$, where the heat flux near the axis would begin to dominate. In addition, we find that the local heat flux is more than twice as high at the location where warm or cold plumes go up or down, than in the plume depleted regions.
\end{abstract}
%%%%%%%%%%%%%%%%%%%%%%%%%%%%%%%%%%%%%%%%%%%%%%%%%%%%%%%%%%%%%%%%%%%%%%%%%%%%%%%%%%%%%%%%%%%%%%%%%%%%%%%

\maketitle

%%%%%%%%%%%%%%%%%%%%%%%%%%%%%%%%%%%%%%%%%%%%%%%%%%%%%%%%%%%%%%%%%%%%%%%%%%%%%%%%%%%%%%%%%%

\section{Introduction}

In Rayleigh-B\'{e}nard (RB) convection a fluid is heated from below and cooled from above. This problem of thermal convection is of the utmost importance from an applied point of view. Examples are thermal convection in the atmosphere, in the oceans and in process technology. For recent reviews of RB convection we refer to Refs.~\cite{lohsermp,lohsearfm}. 

In a cylindrical container, the dynamics of a RB system depends on three control parameters, the Rayleigh number $Ra={g\beta\Delta L^3/{\nu\kappa}}$, the Prandtl number $Pr=\nu/\kappa$, and the aspect ratio $\Gamma=D/L$. Here $g$ is the gravitational acceleration, $\beta$ the isobaric thermal expansion coefficient, $\Delta$ the temperature difference between the top and bottom plates, $\nu$ the kinematic viscosity, $\kappa$ the thermal diffusivity,  and $L$ and $D$ are the height and diameter of the cylinder. The response of the system is expressed by the Nusselt number $Nu$, the dimensionless heat flux~\cite{lohsermp}. 

Previous studies mainly focused on determining the global heat flux as a function of $Ra$ and $Pr$. For water ($Pr=4-6$), in the experimentally available range of $Ra=10^8-10^{11}$, one finds that the global heat transport effectively scales as $Nu \sim Ra^{0.29-0.31}$~\cite{siggia94,gro00,xia02a,ahlers05,ahlers05a,sun05,lohsermp}. The effective exponents for the global heat flux are well described by the unifying theory of Refs.~\cite{gro00,gro01,gro02,gro04}. That theory also made predictions for the scaling exponents of the local heat flux in the center of the cell and at the side wall \cite{gro04}. The reasoning is based on splitting the thermal energy dissipation field into its plume and background contributions; similarly the kinetic energy dissipation is decomposed into its boundary layer and bulk contributions. By doing this Grossmann and Lohse~\cite{gro04} accounted for the various scalings in the $Ra-Pr$ parameter space. They found that the local heat flux has an effective power law dependence on the $Ra$ number, $\overline{j}(r)\propto Ra^{\gamma_{j}(r)}$, and obtained a prediction for the scaling exponent $\gamma_{j}=0.45$ in the bulk (center) and $\gamma_{j}=0.22$ at the plume (side wall) regions. 

In order to understand the heat flux one has to either rely on Eulerian~\cite{shang03} or Lagrangian~\cite{pinton07,schumacher08} measurements where the complex interplay between velocity and temperature can be studied.
Advancements in the experimental techniques made it possible to measure the vertical local velocity $u_z(\mathbf{r}, t)$ and the local temperature $T(\mathbf{r}, t)$ at a given spatial location $\mathbf{r}$ as functions of time. 
This allowed Shang et al.~ \cite{shang03,shang04} to determine the local convective heat flux \cite{footnote}
\begin{equation}
j(\mathbf{r},t) = {{u_z(\mathbf{r},t)[T(\mathbf{r},t)-T_0]} \over {\kappa \Delta}/L},
\label{jequation}
\end{equation}
where $T_0$ is the mean bulk temperature. 
They determined the probability density functions (PDF) of the local heat flux in the axis and side wall regions and showed that the vertical heat flux is highly non-Gaussian and intermittent due to thermal plumes. This work stimulated Ching et al.~\cite{ching04} to theoretically study the problem. They decomposed the velocity field into a part associated with strong temperature fluctuations plus a background and found that, with the definitions they used, the local heat transport associated with the former velocity component near the axis of the cell scaled as $Ra^{1/7}$. Later experiments by Shang et al.~\cite{shang08} revealed that the effective scaling exponent for the local heat flux is about $0.49$ near the cell axis and about $0.24$ near the side wall, which confirms the main results obtained by Grossmann and Lohse \cite{gro04}, but is in disagreement with the model of Ching et al.~\cite{ching04}. 

\begin{figure}[!t]
\vspace*{-2ex}
 \includegraphics[width=0.45\textwidth]{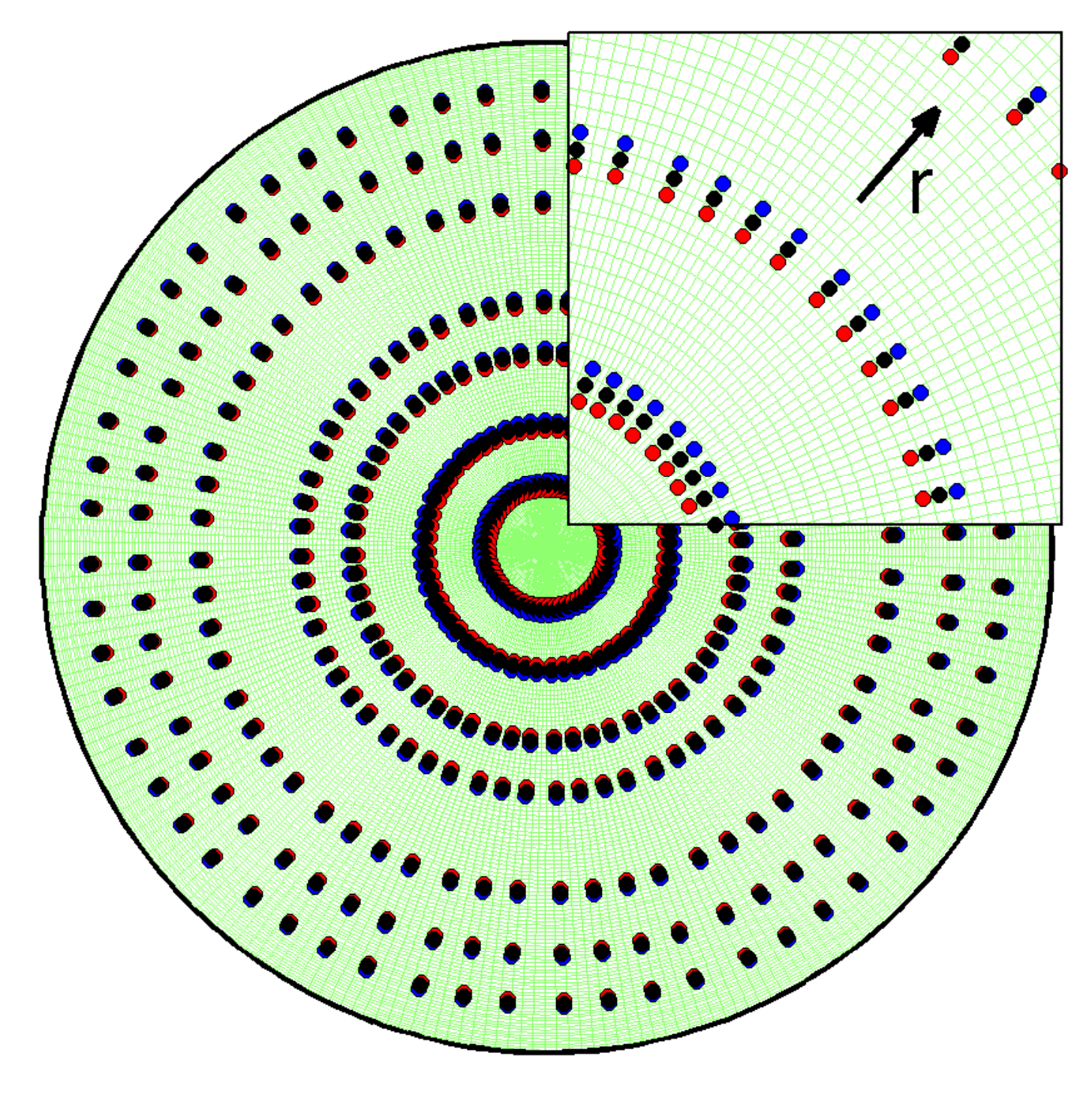} 
  \caption{Computational grid used for $Ra=2\times 10^7$. We position numerical probes uniformly in azimuthal direction in circles at seven radial locations $r/L=0.06$, 0.12, 0.19, 0.24, 0.34, 0.40, 0.45 
  (shown in black). For better statistics we use two more sets of probes (blue, red) with an offset of one grid point. Inset shows a close up view of the probe locations on the grid.}
\label{fig:gridprobe}
\vspace*{-2ex}
\end{figure}

\begin{table*}
\caption{\label{table0} Summary of simulation parameters: The number of grid points used in angular ($N_{\phi}$), radial ($N_r$), and axial directions ($N_z$), volume and time averaged Nusselt number ($Nu$), the number of grid points in the thermal boundary layer ($n_{bl}$), 
convergence of exact relations for $\varepsilon_K$ and $\varepsilon_T$, 
comparing maximum of grid spacing in angular ($\delta\phi_{m}$) and axial ($\delta z_{m}$) directions with the Kolmogorov length scale based on global kinetic energy dissipation rate ($\eta$) and the averaging time considered for the simulations are shown.
The reported times are always measured in the units of free fall time and the lengths in terms of cylinder height. 
For statistical averages we discarded the initial 140 dimensionless times, to prevent transient effects in the results.  
}
   \begin{center}
  { \setlength{\tabcolsep}{6pt}
    \begin{tabular}{ | c | c | c | c | c | c | c | c |}
    \hline 
       
    $ Ra$		&  $N_{\phi}\times N_{r}\times N_{z}$	& $Nu$	& $n_{bl}$ & ${\varepsilon_K\over {{\nu^3\over L^4} {Ra\over Pr^2} (Nu-1)}}$ 	&  ${\varepsilon_T\over {\kappa{\Delta^2\over L^2} Nu}}$ 		
    &		$\delta \phi_{m}$, $\delta z_{m}$, $\eta$ & in $L/U$ \\
    & & & & & & ($\times 10^2$)		&   \\  \hline
                                                                               
    $2\times 10^6$		& $193\times 49\times 129$		& 10.93	& 19	& 1.008	& 0.968	& 2.26, 1.45, 3.41	 & 4200    \\ 
     $1\times 10^7$		& $257\times 65\times 193$		& 16.58	& 18	& 1.010	& 0.971	& 2.44, 0.73, 2.04	 & 3700  \\ 
    $2\times 10^7$		& $257\times 65\times 193$		& 20.71	& 18	& 1.007	& 0.843	& 2.44, 0.73, 1.62	 & 3700  \\ 
    $6\times 10^7$		& $321\times 97\times 239$		& 28.48	& 14	& 0.989	& 0.923	& 1.96, 0.97, 1.13	 & 3000  \\ 
    $1\times 10^8$		& $385\times 129\times 257$		& 33.25	& 15	& 0.997	& 0.955	& 1.62, 0.73, 0.95	 & 2800  \\
    $2\times 10^8$		& $385\times 129\times 257$		& 40.87	& 13	& 1.002	& 0.937	& 1.62, 0.73, 0.76	 & 2700  \\
    $5\times 10^8$		& $513\times 161\times 321$		& 52.80	& 13	& 1.005	& 0.947	& 1.22, 0.51, 0.56	 & 1900  \\
    $2\times 10^9$		& $769\times 193\times 385$		& 80.34	& 11	& 0.992	& 0.959	& 0.82, 0.34, 0.26	 & 2040  \\  \hline
    
  \end{tabular}  
 }     
\end{center}
\label{table0}
\vspace*{-2ex}
\end{table*}    

In experiments it is quite difficult to measure the heat flux at each spatial point in the cell due to problems with measurement techniques and the presence of inherent noise levels. The measurements at just one or two points may not be enough to understand the complex dynamics involved in the convection process. The present paper offers numerical results which complement the work initiated by Shang et al.~\cite{shang03,shang08}. We provide information on the heat flux at one quarter, one half, and three quarters of the cell height for several radial positions $r$, not only near the axis 
and the side wall. This information allows us to understand the two limits in the unifying theory~\cite{gro04} on the effective scaling exponents, one valid in the bulk (the central region of the cell) and the other valid in the plume region. As a result we clearly see persistence of the inhomogeneous nature of the flow in the radial direction which leads to different scaling exponents. 

In experiments the local velocity is measured by LDV/PIV techniques at a spatial position which slightly differs from the location of the local temperature measurement. This spatial misalignment may possibly affect the results. In numerical simulations, in contrast, one has all the information on the flow field and thus the local heat flux can be calculated from the velocity and temperature measurements at exactly the same position. 
Finally we comment on the local heat flux distribution with respect to the large scale circulation and the ultimate regime mentioned in Refs.~\cite{gro01,toschi03,gro04} and provide data to illuminate the differences between the measurements of Shang et al.~\cite{shang08} and the earlier Ching et al.~\cite{ching04} claims. 

In simulations the $Ra$ number range and the duration available for time-averaging are more limited than in experiments. In order to mitigate the latter shortcoming we have limited the $Ra$ number of our simulations. In the next section we briefly describe the numerical procedure before discussing our results on the local heat flux and the velocity and temperature fluctuations.

\section{Numerical method}
  
We performed direct numerical simulations for a Boussinesq fluid in a unit aspect ratio ($\Gamma=1$) cylinder with constant temperatures applied at  the top and bottom plates and an adiabatic side wall. The fluid simulated in our calculations is water at 32 $^o$C ($Pr=5.2$) for 
$2\times10^6 \leq Ra \leq 2\times 10^9$. The governing equations for momentum, energy and mass conservation in dimensionless form are given by \cite{verzicco03} 
\begin{eqnarray}
{D\mathbf{u}\over Dt} & = & -\nabla p +  \theta \mathbf{\hat z}+ {\left({Pr \over Ra}\right)^{1/2}} \nabla^2 \mathbf{u},  \\
{D\theta \over Dt} & = & {1\over(PrRa)^{1/2}} \nabla^2 \theta, \quad  \nabla\cdot\mathbf{u} = 0. 
\end{eqnarray}
Here the dimensionless variables are the velocity $\mathbf{u}$, temperature $\theta$ and pressure $p$ (minus the hydrostatic contribution).
The material derivative is denoted by ${D/Dt}$. The unit vector $\hat{\mathbf{z}}$ is in the direction opposite to gravity.  
The physical variables such as length and velocity are made non-dimensional by the cylinder height ($L$), 
and the free-fall velocity $U=\sqrt{g\beta\Delta L}$. Constant dimensionless temperatures of 1 and 0 are applied at the bottom and top plate, respectively. 

The governing equations are solved on a staggered grid with second-order accuracy in space and time. 
For the time advancement a third order Runge-Kutta scheme is used. This method is stable for a CFL number up to $\sqrt{3}$, and here we have limited it to $1.2$ by adjusting the time step according to the maximum velocity before executing each time step \cite{orlandi2001}. 
More details about the numerical method can be found in Refs.~\cite{verz96,verz97a,verzicco03}. For the spatial resolution we followed the criteria set by Stevens et al.~\cite{richard10}. 

A summary of the simulation parameters is shown in Table~\ref{table0}. The first and second columns are $Ra$ and the number of grid nodes. 
The volume- and time-averaged global heat transport, i.e. the Nusselt number 
$Nu=1+\sqrt{Ra Pr} \left[ \overline{{\langle{u_z \theta}\rangle_V}} \right]$, is shown 
in the third column; here the overline denotes time averages and the angular brackets $\langle\cdot \rangle_V$ volume averages. 
The number of grid points used to resolve the thermal boundary layers is shown in the fourth column. In a RB cell with no slip velocity condition on the walls the dimensional thermal  energy dissipation rate, 
 $\varepsilon_T = \kappa\Delta^2 L^{-2} {\overline{\langle{ |\nabla \theta|^2}\rangle_V}}$ and kinetic energy dissipation rate  
 $\varepsilon_K = \nu^3 L^{-4} Pr^{-1} Ra {\overline{\langle{ |\nabla \mathbf{u}|^2}\rangle_V}}$ 
satisfy exact relationships with $Nu$ (see Refs.~\cite{siggia90,lohsermp}), namely $\varepsilon_T =  \kappa\Delta^2 L^{-2} Nu$ and $\varepsilon_K = \nu^3 L^{-4} Pr^{-2} Ra (Nu-1)$. 
In order to validate our simulations the obtained energy dissipation rates are compared with $Nu$ in Table~\ref{table0}. 
These ratios are near one, which proves the adequacy of the grid resolution.
In the seventh column the largest grid spacings in the azimuthal and axial directions are compared with the Kolmogorov length.
In the last column the total time used for the statistical averages is shown in free fall time units ($L/U$). The total computational time 
was around $2.2\times 10^5$ CPU hours on a Power6 computer.

\begin{figure}[!t]
\vspace*{-2ex}
  \includegraphics[width=0.4\textwidth]{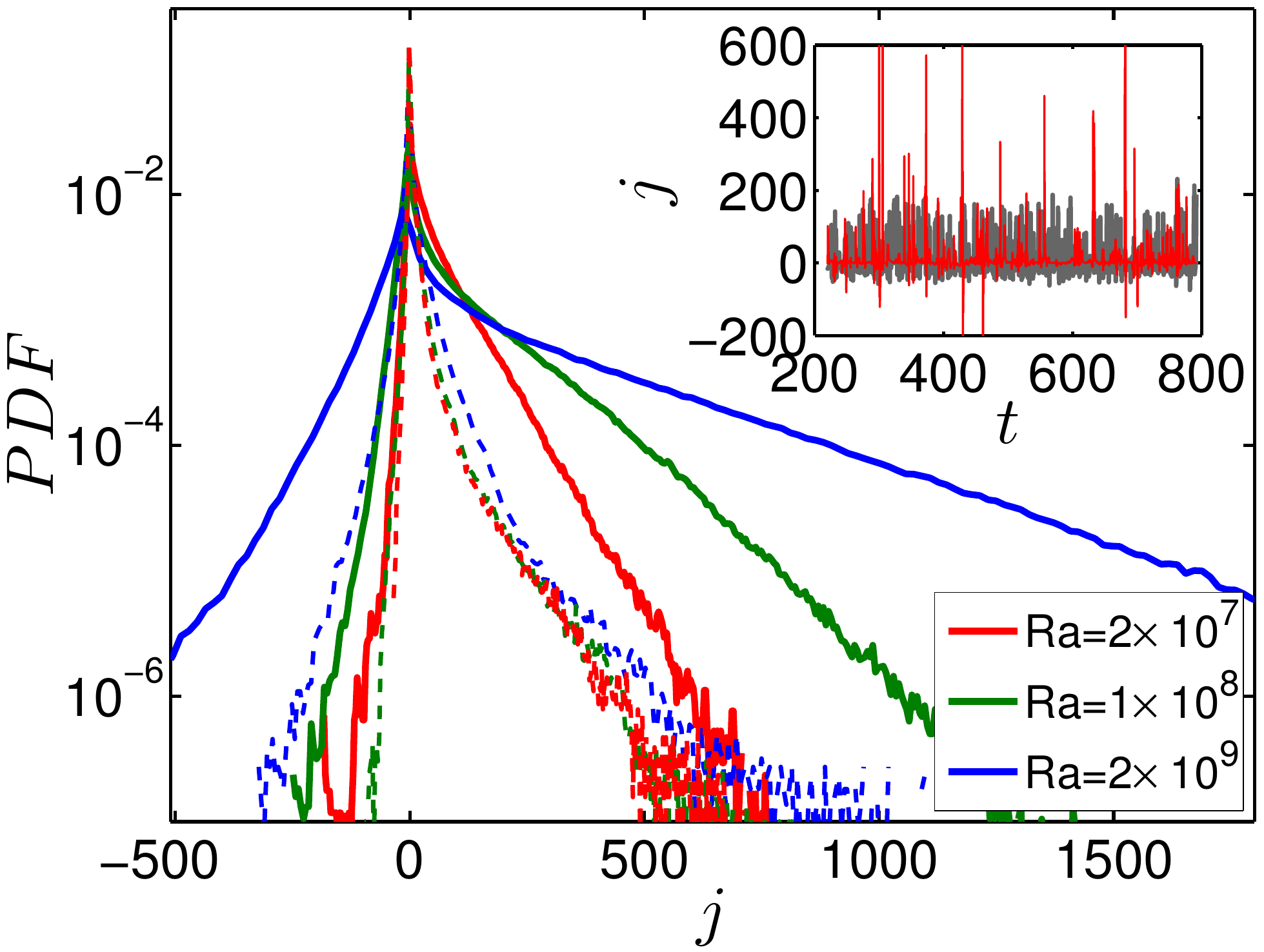} \\
 %   \vspace*{-2ex}
  \caption{PDFs of the local instantaneous heat flux $j$ near the side wall at $r/L=0.45$ (solid, thick lines) and near the axis $r/L=0.06$ (dashed, thin lines) at mid-height for $Ra=2\times 10^7$ (red), $1\times 10^8$ (green), $2\times 10^9$ (blue). The inset shows time series for the heat flux near the side wall (thin, red) and in the center (thick, gray) for $Ra=1\times 10^8$.}
\label{time_pdf}
\vspace*{-2ex}
\end{figure}

\begin{figure}[!t]
\vspace*{-2ex}
\includegraphics[width=0.4\textwidth]{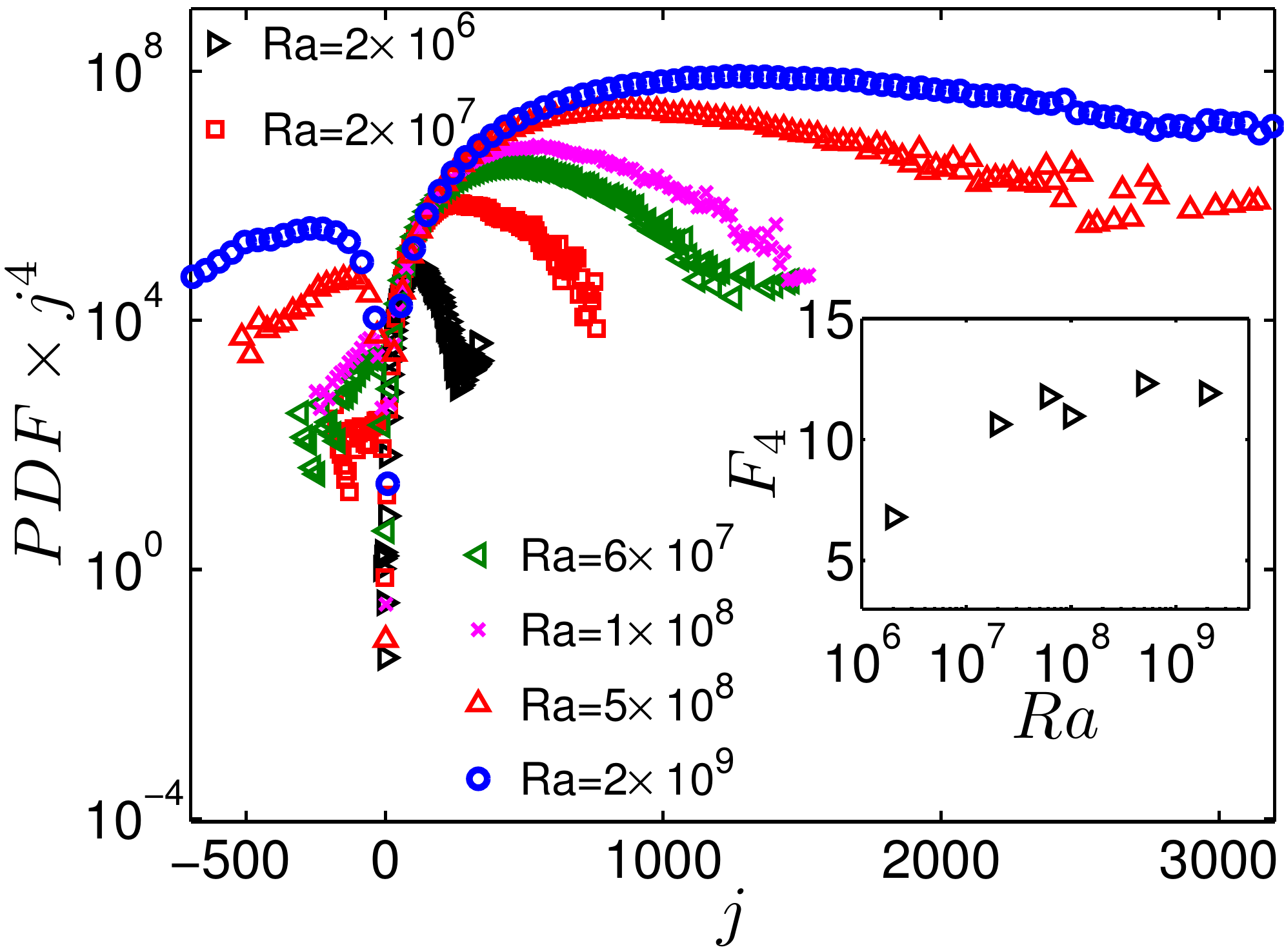} \\
   \caption{${j^4} \times PDF(j)$ vs. $j$ is shown for different $Ra$ near the side wall region showing convergence of the flatness at that location. Note that the y-axis is given in log-scale, in contrast to the figures shown in the convergence test by Belin et al.~\cite{belin96}, where $y^4 \times PDF(y)$ is given on a linear scale. The inset shows that the flatness $F_4$ increases as a function of $Ra$.}
\label{foum}
\vspace*{-2ex}
\end{figure}

We placed 1880 ``numerical probes'' at different radial locations on three different horizontal planes 
(at $z/L$=0.25, $z/L$=0.50 and $z/L$=0.75), to obtain point-wise data on the temperature and vertical velocity in order to 
calculate the local heat flux according to equation (\ref{jequation}). In each horizontal plane a number of azimuthally nearly equally spaced probes were placed on seven circles with radii $r/L=0.06$, 0.12, 0.19, 0.24, 0.34, 0.40 and 0.45, see Figure~\ref{fig:gridprobe}. On each circle we distributed 60 probes at $z/L=0.50$ and 20 probes at $z/L=0.25$, 0.75. 
In addition, on the mid-plane, each circle was complemented by two other circles, one inside and one outside, spaced by one radial mesh length as shown in the upper right corner of Figure~\ref{fig:gridprobe}. In total, we have information from 
60 (azimuthal) $\times$ 7 (radial) $\times$ 3 (sets)$=1260$ probes at mid-height plane. 
For the planes at $z/L=0.25$ and at $z/L=0.75$ we have information from $20\times 7=140$ probes. 

As a further check on the consistency of the calculations we compare the time- and area-averaged heat flux as monitored by all the mid-plane probes 
mentioned before complemented by 340 additional ones uniformly spaced in the angular direction and non-uniformly in the radial direction. 
For $Ra=5\times10^8$ we found a global average Nusselt number of 51.58 while from the $r$-averaged heat flux from the probes at mid-height we found $Nu=50.83$. Similarly, for $Ra=2\times10^9$, the two results were 78.80 and 76.41. 
In view of the discretization error introduced by the sparseness of the numerical probes (5\% of the actual computational cells on the mid-plane), this can be considered as a good agreement. 

\begin{figure}[!t]
    \subfigure{\includegraphics[width=0.4\textwidth]{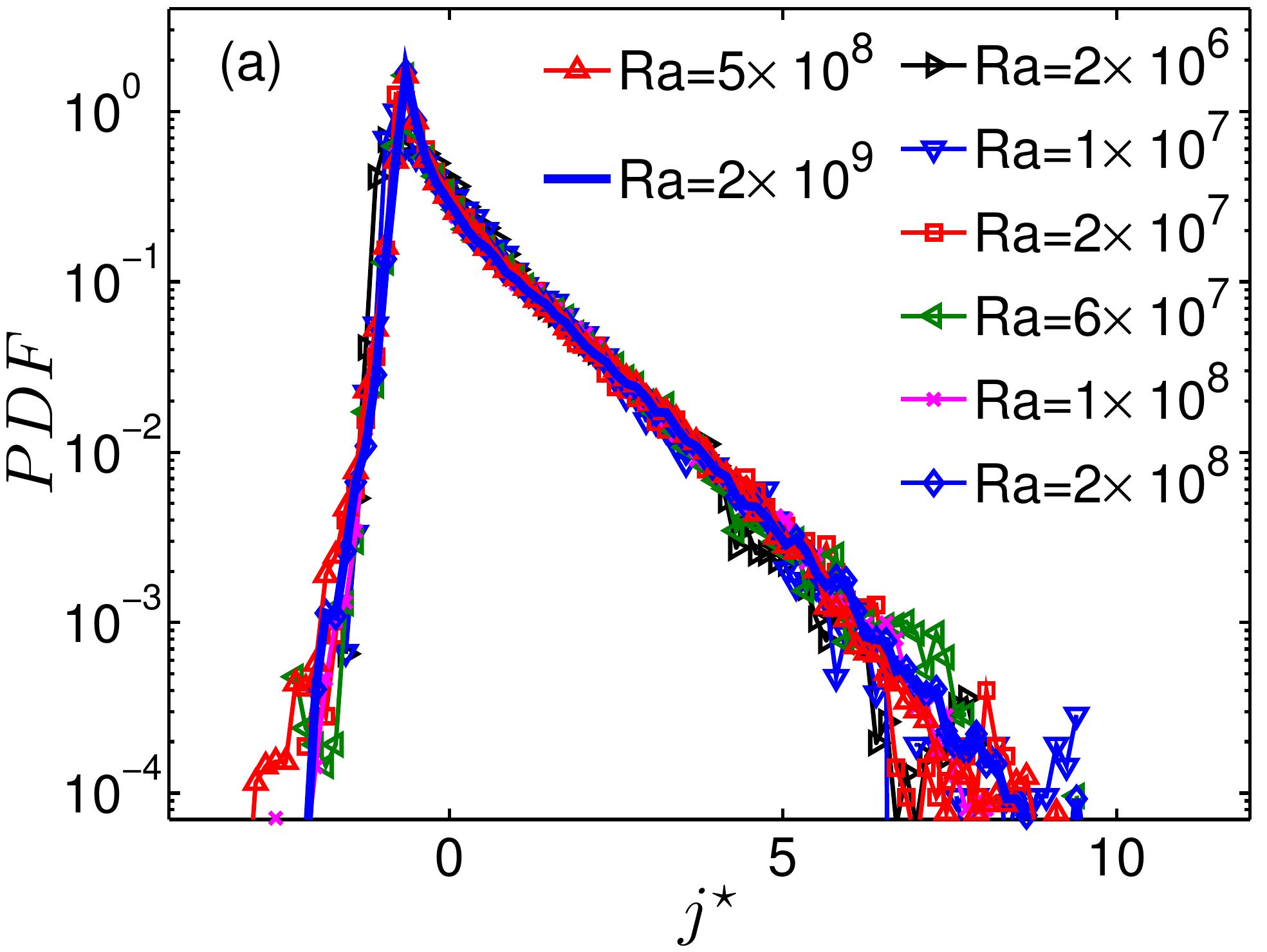}} \\
    \subfigure{\includegraphics[width=0.4\textwidth]{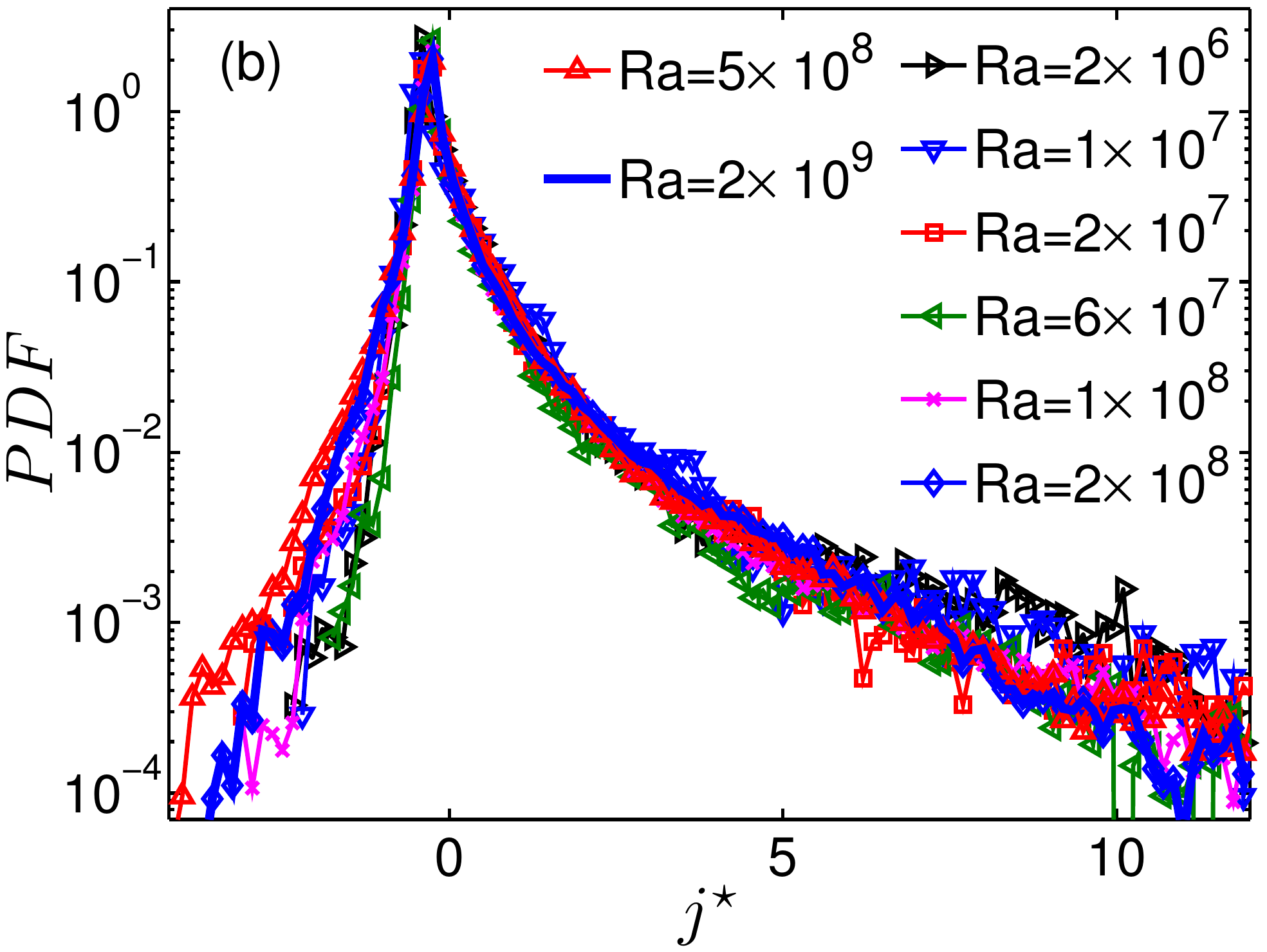}}
    \vspace*{-2ex}
  \caption{Centered PDFs of the normalized heat flux ${j^{\star}}=(j-\overline{j})/j_{rms}$ on the mid-plane (a) near the side wall $r/L=0.45$, and 
  (b) near the axis $r/L=0.06$.
  }
\label{pdf_jz}
\vspace*{-2ex}
\end{figure}

\section{Results}

\subsection{Local heat flux}

In Figure \ref{time_pdf} we show PDFs for the local instantaneous heat flux $j$ on the mid-plane, at two different radial positions, 
one near the cell axis ($r/L=0.06$) and one near the side wall ($r/L=0.45$). Note that the latter is outside the kinematic boundary layer (BL) which
at this Prandtl number has a thickness of $\lambda_u/L \approx 3.6 Ra^{-0.26\pm 0.03} $ \cite{qiu98b}. According to this scaling law
we have $\lambda_u/L \approx 0.05$ for $Ra=10^7$ and $\lambda_u/L \approx 0.016$ for $Ra=10^9$. 
From  our numerical calculations of the kinetic side wall BL
thickness, as identified by the location of the maximal velocity fluctuations, we get even slightly smaller values. Obviously, there is no thermal BL at the side walls due to the adiabatic boundary conditions.

\begin{figure}[!t]
\vspace*{-2ex}
    \subfigure{\includegraphics[width=0.4\textwidth]{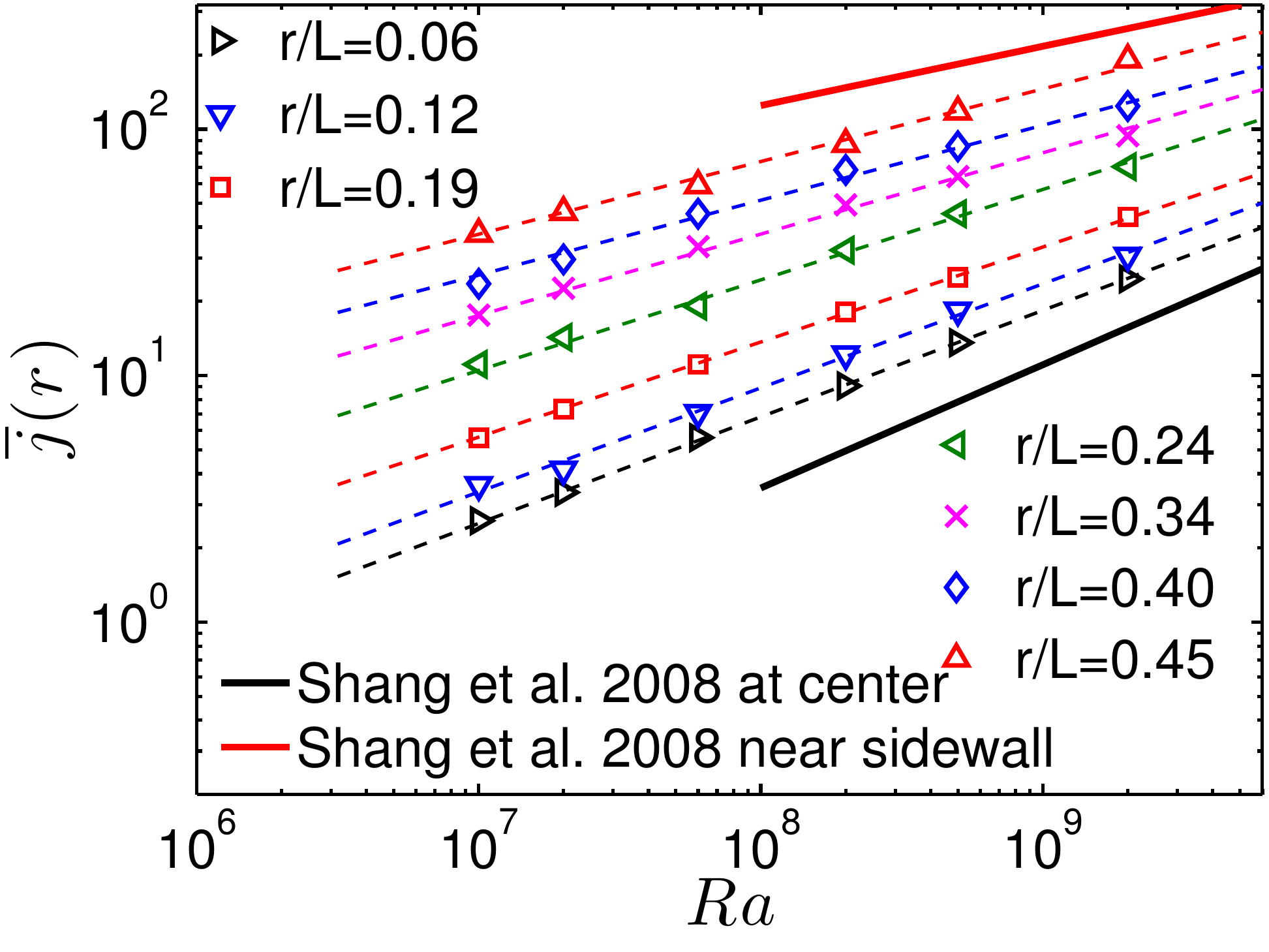}}
    \subfigure{\includegraphics[width=0.4\textwidth]{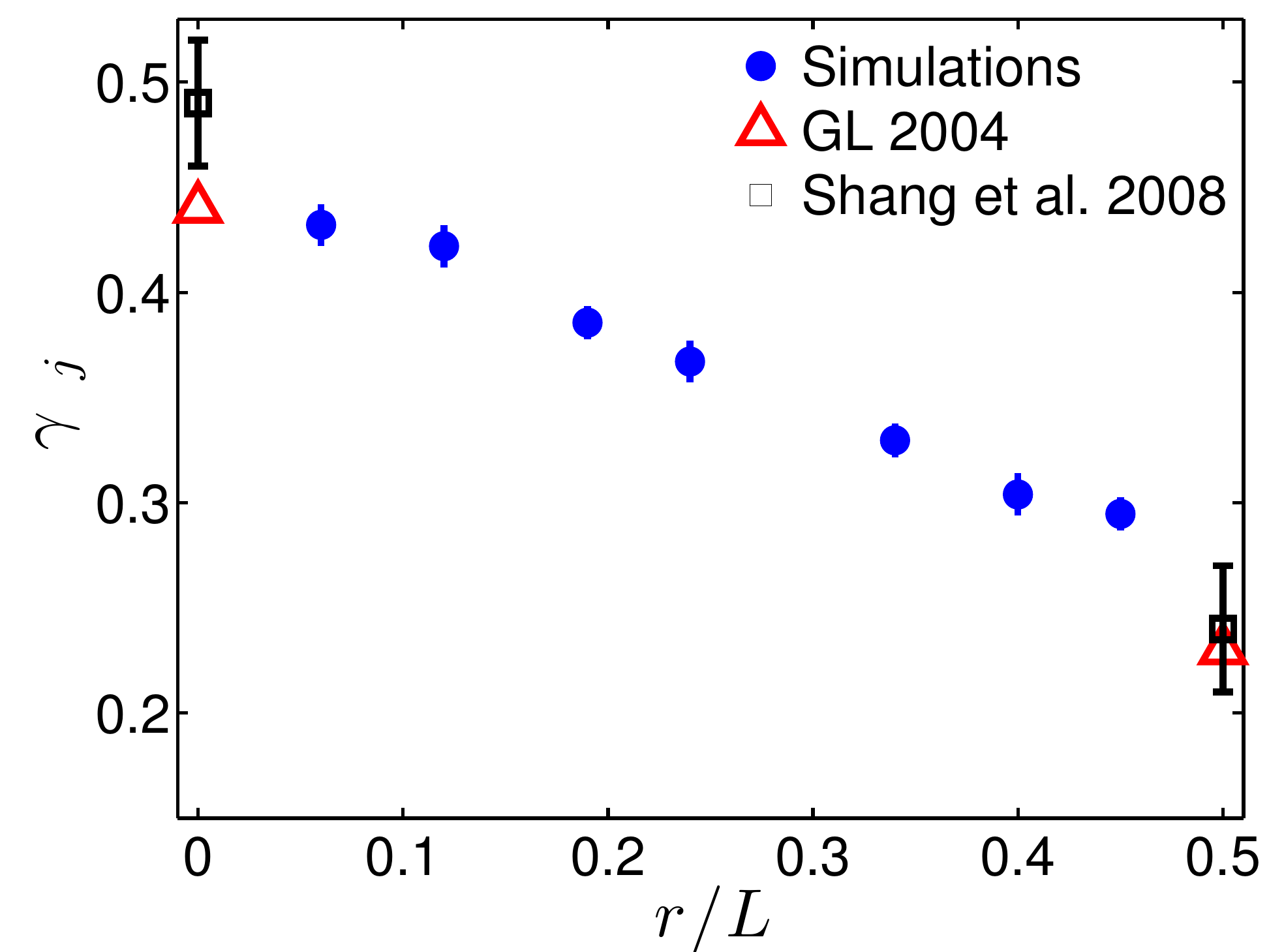}}
    \vspace*{-2ex}
  \caption{(a): Open symbols indicate the numerical results for the local heat flux averaged over time and angular position as function of $Ra$ at different radial positions $r/L$. The solid lines show the experimental data of Shang et al. \cite{shang08}. (b): The scaling exponent for the time- and angle-averaged heat flux as function of the radial position $r/L$ for the simulations, experiment~\cite{shang08}, and theory~\cite{gro04}.}
\label{jz_vs_ra}
\vspace*{-2ex}
\end{figure}

\begin{figure}[!t]   
\vspace*{-2ex} 
    \subfigure{\includegraphics[width=0.4\textwidth]{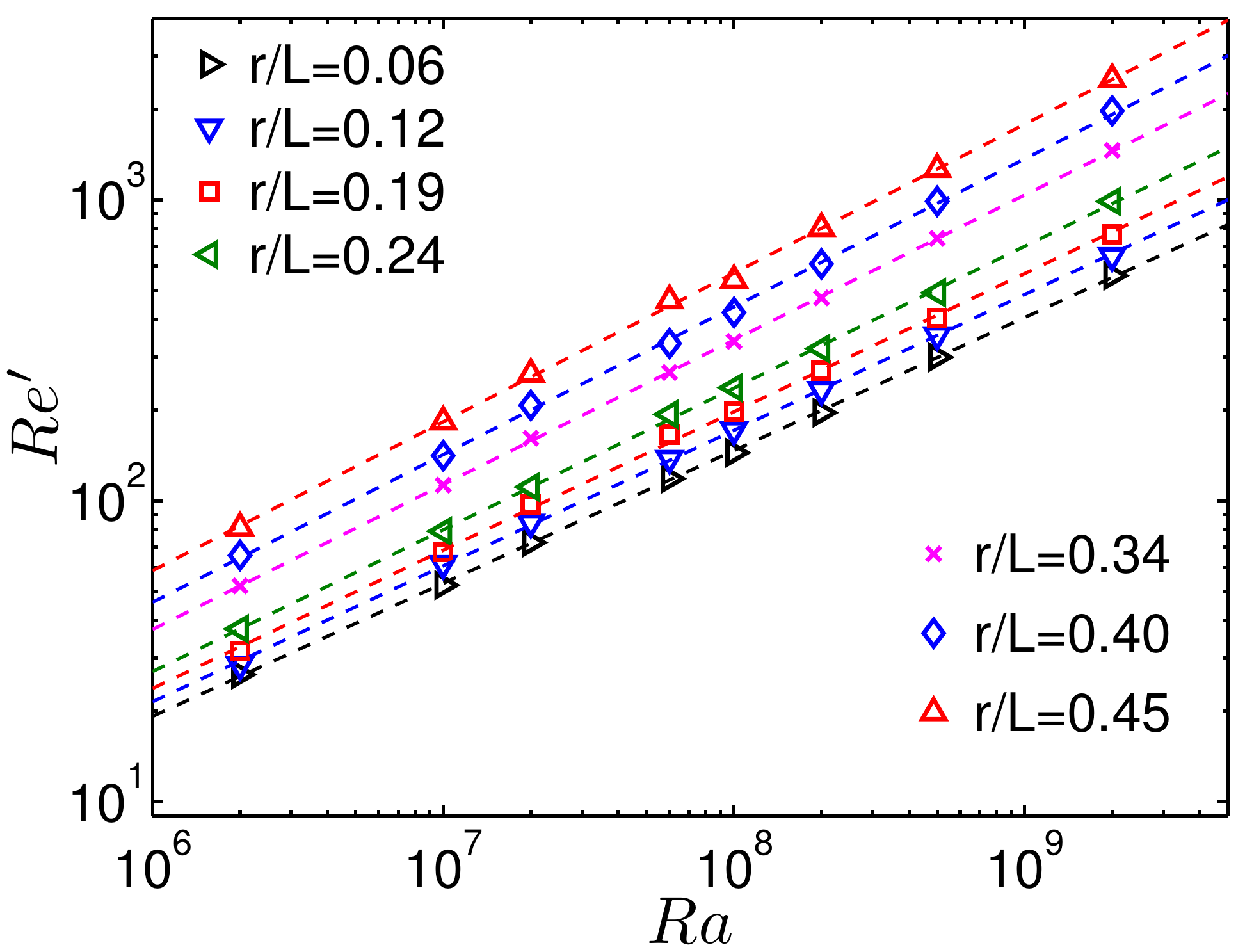}}
    \subfigure{\includegraphics[width=0.4\textwidth]{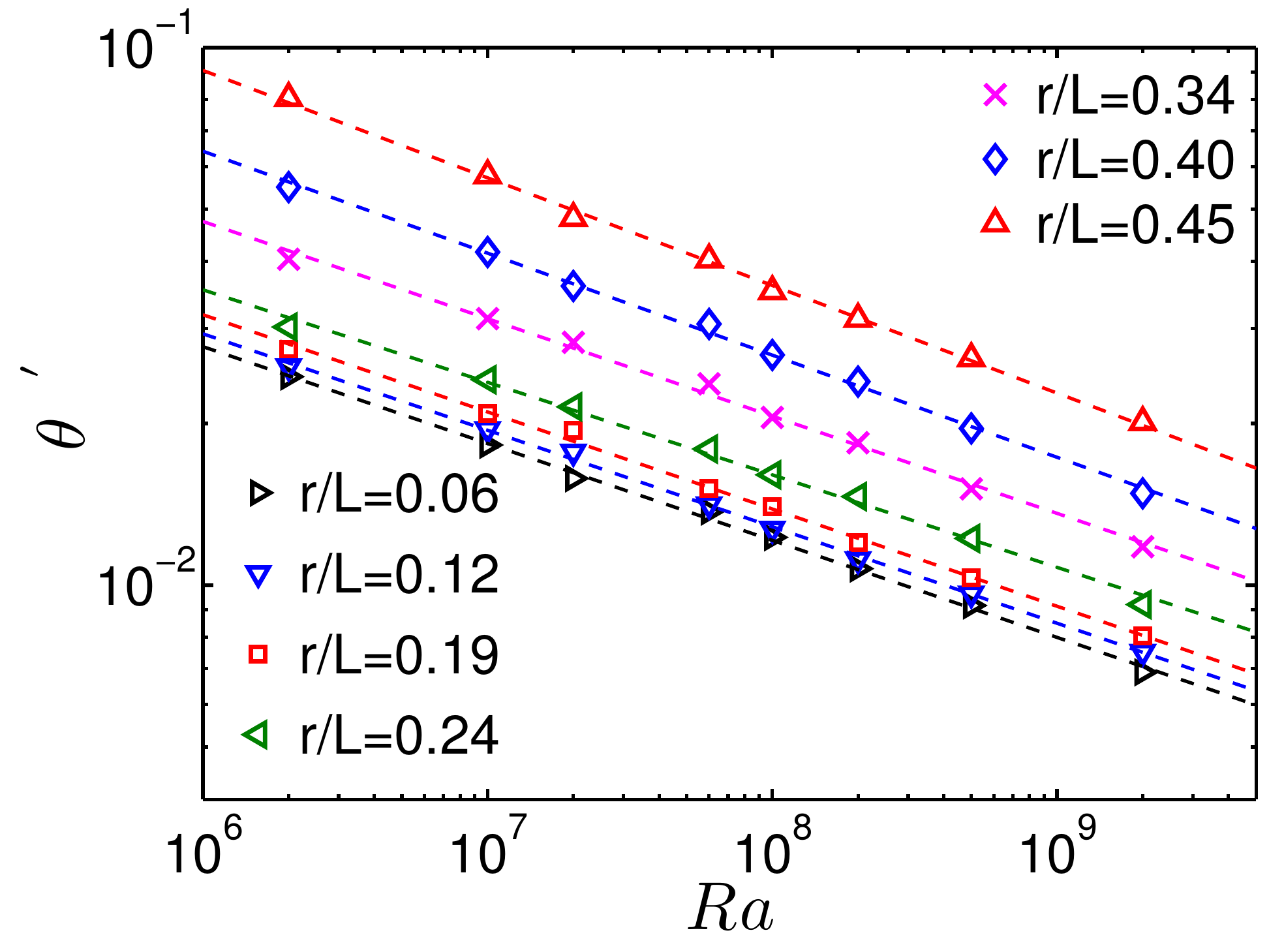}}
    \vspace*{-2ex}
 \caption{Panel (a) shows the Reynolds number based on the rms velocity, $Re^{\prime}$, and panel (b) the normalized rms temperature $\theta^{\prime}$ as function of $Ra$; both for different radial positions.  
}
\label{reT_vs_ra}
\vspace*{-2ex}
\end{figure}

The first striking observation from Figure~\ref{time_pdf} is that the absolute value of the heat flux is much larger at the side walls (solid lines) 
as compared to the center (dashed lines). This is consistent with the expectation that most of the heat is transported by the large-scale convection roll 
and of course it is well known \cite{shang08}. 
In addition, we observe that the PDF close to the side walls has a positive skewness due to rising and falling of 
the warm and cold plumes. We observe some events with a heat flux as high as fifteen times the average. The PDF of the local heat flux in the 
cell center also has a positive skewness, which indicates that plumes can travel through this region as well. The inset of Figure  \ref{time_pdf} shows $j$ as a function of the dimensionless time $t$ close to the cell center ($r/L$=0.06) and to the side wall ($r/L$=0.45) for $Ra=10^8$ and again reflects  the presence of much stronger heat transport events near the side wall than in the center. 

One of the main features of turbulence is the small scale intermittency that is measured as departure from a Gaussian character of the PDF, mainly the tails and the peakedness. This can be quantified by calculating the flatness $F_4$ of the PDF. For strongly intermittent signals however the integral 
of ${j^4} \times PDF(j)$ defining the flatness may not converge. To examine this issue we calculate the angular average of this quantity at $r/L=0.06$ and at $r/L=0.45$ on the mid-plane. While at the side wall this quantity decays for large $\mid j \mid$ sufficiently fast, see Figure~\ref{foum} (this feature is more evident when data is plotted on a linear rather than log-scale), and thus permits the calculation of the flatness (showing strong intermittency, $F_4 \approx 11$, inset of Figure~\ref{foum}), in the center the intermittency is so strong that no convergence for the flatness can be achieved. 

By rescaling the heat flux $j$ with its standard deviation $j_{rms}$, the zero-mean PDF for the normalized  heat flux $j^{\star} \equiv(j-\overline{j})/j_{rms}$ shows universality near the side wall (see Figure~\ref{pdf_jz}a). The tails for the rescaled PDFs are shorter at the side walls compared to those at the center. This indicates relatively fewer plumes carrying a large heat flux at the side walls in contrast to relatively more plumes carrying a smaller heat flux at the center, and again underlines the extremely strong intermittency of the heat flux in the center (see Figure~\ref{pdf_jz}b). 

Figure~\ref{jz_vs_ra}a shows the time- and angularly-averaged heat flux as a function of $Ra$ at different radial positions on the mid-plane. 
Figure~\ref{jz_vs_ra}b shows that the corresponding scaling exponent $\gamma_{j}$ as a function of the radial position decreases monotonically from $0.43$ near the axis to $0.29$ close to the side wall. The measurements of Shang et al.~\cite{shang08} and the theoretical analysis of Grossmann and Lohse~\cite{gro04} only made statements on the values close to the side wall and to the axis; these are well confirmed by the present results. In the $Ra$ range considered here the heat transport near the side wall is an order of magnitude larger than in the center. An extrapolation of the power law fits $\overline{j}=0.0025Ra^{0.43\pm0.01}$ obtained near the center and $\overline{j}=0.3236Ra^{0.29\pm0.01}$ valid near the wall, shows that these become equal for $Ra \approx 10^{15}$. This value is consistent with the prediction of the unifying theory \cite{gro01}. 
Recent experiments~\cite{ahlers11a} suggests the occurrence of this feature in the range $10^{13}\le Ra \le 5\times 10^{14}$, whereas Shang et al.~\cite{shang08} suggested that it happens at $Ra\approx 10^{14}$, based on their extrapolation.
\begin{figure}[!t]
\vspace*{2ex}
\includegraphics[width=0.45\textwidth]{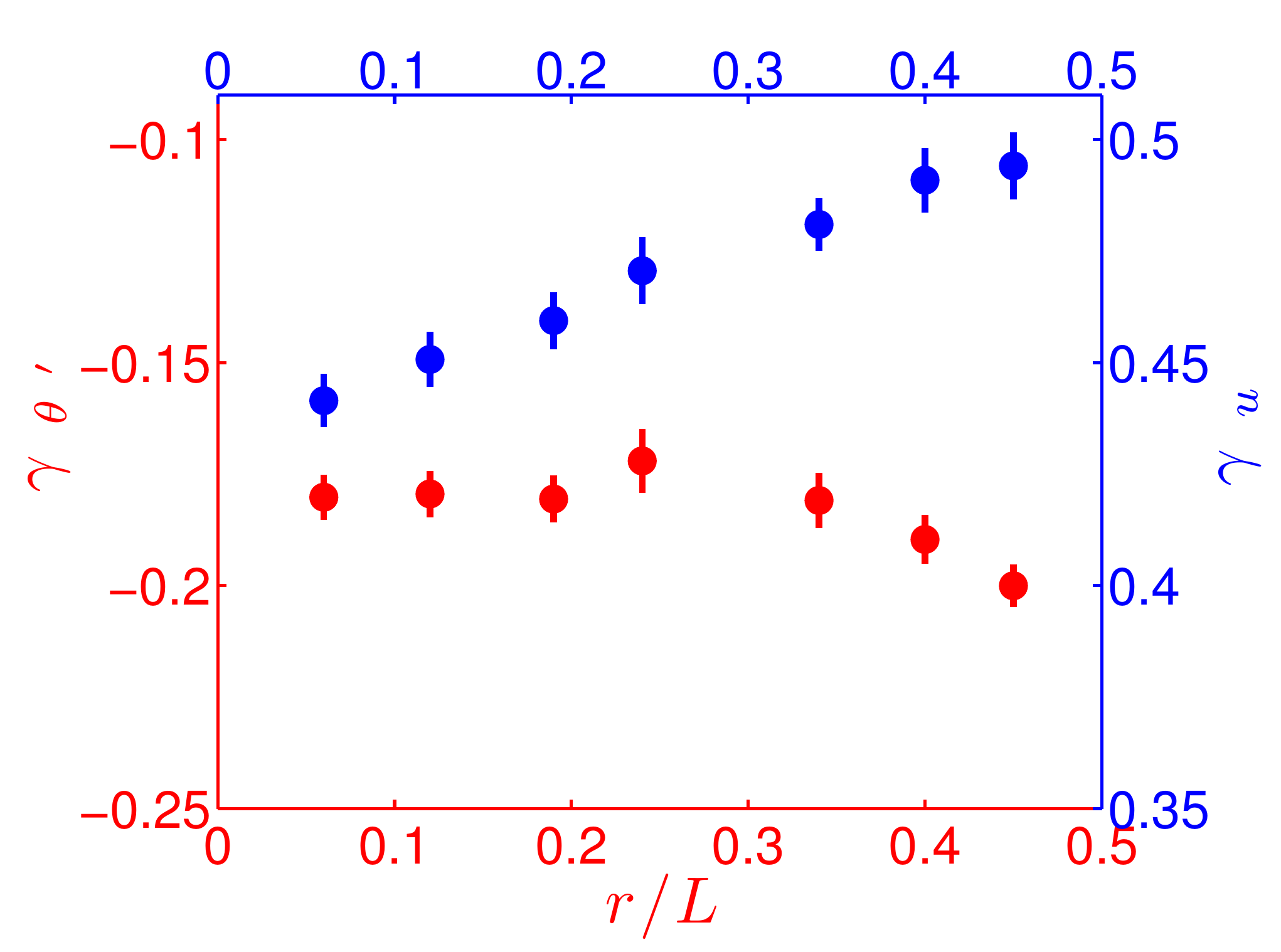} 
 \caption{The variations in the velocity and temperature scaling exponents with $Ra$ are shown as functions of the radial position.
}
\label{scale_uT}
\vspace*{-2ex}
\end{figure} 

\begin{figure}
\vspace*{2ex}
\includegraphics[width=0.4\textwidth]{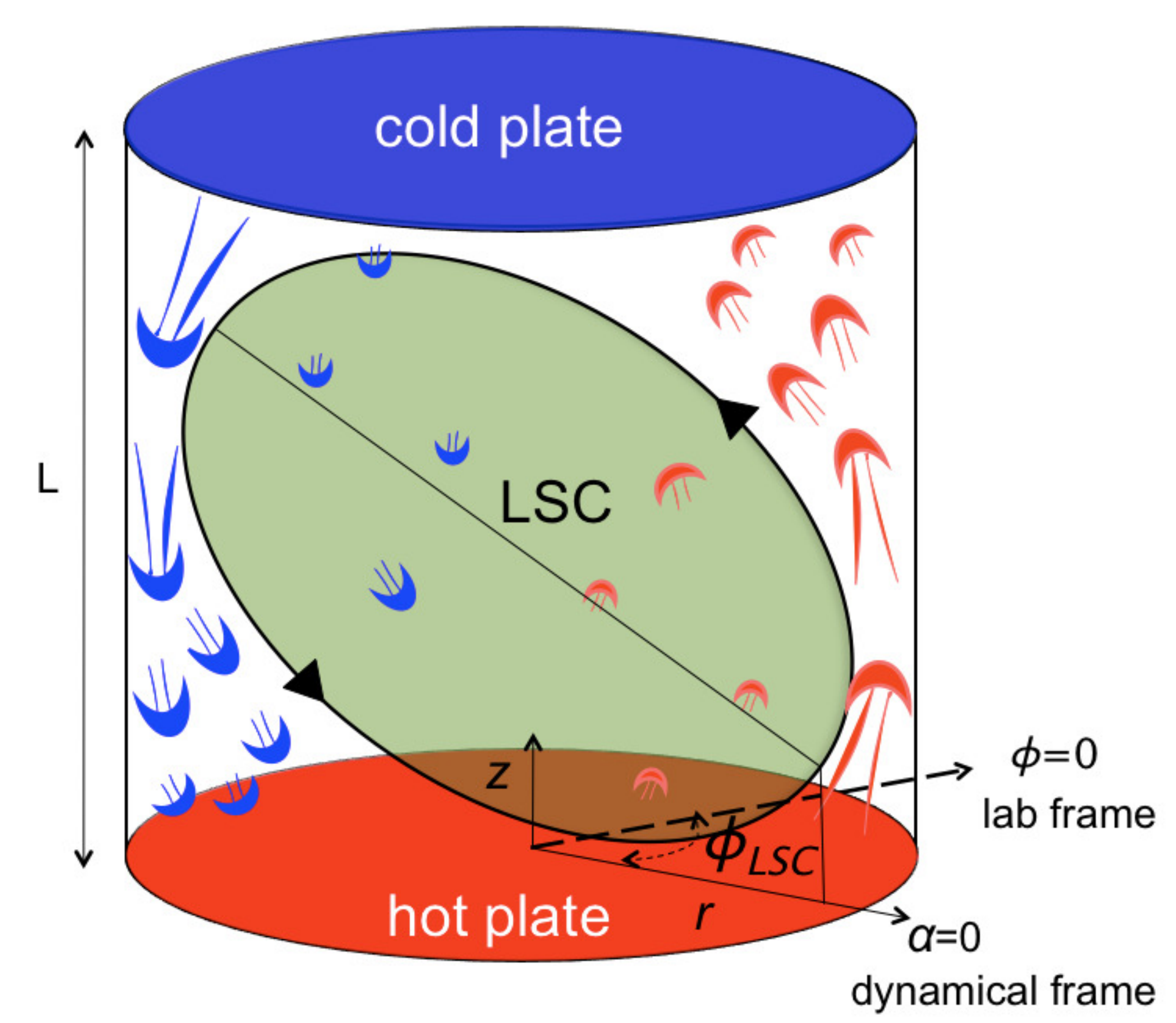} 
 \caption{Sketch of the LSC in a cylindrical RB cell. The warm plumes go upwards on the right side, defined as $\alpha=0$, and go downwards on the left side of the cell, defined as $\alpha=\pi$. Due to the shape of the LSC the warm uprising fluid (cold down flowing fluid) is close to the side wall at $z/L=0.25$ and $z/L=0.50$ ($z/L=0.75$ and $z/L=0.50$).
}
\label{fig:plumes_sketch}
\vspace*{-2ex}
\end{figure} 

\begin{table*}
\caption{\label{table1} Summary of the velocity and temperature scaling exponents $\gamma_u$ and $\gamma_{\theta}$ reported in several experimental (E) and theoretical (T) studies. The experiments mentioned below have been carried out in cylindrical cells, unless stated otherwise. }
    \begin{center}
{\footnotesize
    \begin{tabular}{ | c | c | c | c | c | c | c | c |}
    \hline
			  						& $Ra$ 						& $Pr$ 			& $\Gamma$ 		& $\gamma_u$ (center) 		& $\gamma_{\theta}$ (center)	& $\gamma_u$ (side wall)	& $\gamma_{\theta}$ (side wall)	\\  \hline                                                                               
     Castaing et al.\ \cite{libchaber89} (E)	 	& $4\times 10^7-6\times 10^{12}$ 	& $0.65-1.5$ 		& $1$  			&  $0.491\pm 0.002$		& $-0.147\pm 0.005$	       	& --					&  --						\\ \hline
     Castaing et al.\ \cite{libchaber89} (T)	 	& $4\times 10^7-6\times 10^{12}$ 	& $0.65-1.5$ 		& $1$			&  $3/7$					& $-1/7$				       	& --					& --						\\ \hline
     Sano et al.\ \cite{sano89} (E) 			& $10^8-10^{10}$				& $0.64-1.4$ 		& $1$ 			& --						& --						& $0.485 \pm 0.005$ 	& --   						\\ \hline
     Takeshita et al.\ \cite{takeshita96} (E)		& $10^6-10^{8}$ 				& $0.024$ 		& $1$ 			& --						& --						& $0.46 \pm 0.02$ 		& --   						\\ \hline      
     Ashkenazi et al.\ \cite{ash99,ash99b} (E)	& $10^{11}-5\times 10^{14}$ 		& $27-190$ 		& $1$ (square)  	& --						& --						& $0.43 \pm 0.02$		& --   						\\ \hline   
     Chavanne et al.\ \cite{chavanne01} (E) 	& $10^7-6\times 10^{12}$ 		& $0.7-4$	 		& $0.5$ 			& --						& --						& $0.49$ 				& --    					\\ \hline   
     Daya and Ecke \cite{daya01} (E)			& $2\times 10^8-4\times 10^{9}$ 	& $5.46$ 			& $0.79$  			& $0.5 \pm 0.03$  			& $-0.10 \pm 0.02$     		& --					& --						\\ \hline
     Niemela et al.\ \cite{niemela01} (E) 		& $15\times10^6 - 10^{13}$		& $0.7$			& $1$  			& --						& --						& $0.5$				& --   						\\ \hline     
     Qiu et al.\ \cite{qiu02,qiu04} (E) 			& $10^8-10^{10}$ 				& $5.4-5.5$ 		& $1$ 			& $0.55$ 					& --   						& $0.46$				& --						\\ \hline 
     Lam et al.\ \cite{lam02}	(E)				& $10^6-10^{11}$ 				& $6-1027$	 	& $0.5-4.4$ 		& --	 					& --						& $0.495$ (bottom)		& --						\\ \hline
     Grossman and Lohse \cite{gro04} (T)	& $10^6-10^{14}$ 				& $\sim 0.1-10$ 	& $1$ 			& $0.34$ 					& $-0.11$ to $-0.16$     		& --					& $-0.09$ to $-0.11$			\\ \hline
     Shang et al.\ \cite{shang08} (E)			& $10^8-10^{10}$ 				& $4.4$ 			& $1$ 			& $0.49 \pm 0.03$  			& $-0.14 \pm 0.03$     		& $0.46 \pm 0.03$		& $-0.24 \pm 0.03$			\\ \hline
     Present work 						& $2\times10^6-2\times 10^9$ 		& $5.2$ 			& $1$ 			& $0.44 \pm 0.01$  			& $-0.18 \pm 0.01$  			& $0.49 \pm 0.01$		& $-0.20 \pm 0.01$			\\ \hline
  \end{tabular}  
 }     
\end{center}
\label{table1}
\vspace*{-2ex}
\end{table*}

\subsection{Local fluctuations}

In this section, we determine the scaling with $Ra$ of the velocity and temperature fluctuations with respect to their global mean at different radial locations. Since the global mean of the vertical velocity vanishes, to characterize the velocity fluctuations $u_{rms}$ it is sufficient to take the root mean square of the fluid velocity at each position and average it in the azimuthal direction. Figure~\ref{reT_vs_ra}a shows $Re^{\prime}\equiv \sqrt{(Ra/Pr)}u_{rms}$ (equivalent to $U_{dim} L/\nu$ with $U_{dim}$ the dimensional rms velocity) as a function of $Ra$ on the mid-plane. 

For the normalized temperature fluctuations we take the root mean square of $\theta'=\theta -1/2$. Figure \ref{reT_vs_ra}b shows the results for this quantity vs. $Ra$ at different radial positions on the mid-plane. 
At all radial positions, fluctuations of both velocity and temperature exhibit a power law dependence on $Ra$ proportional to $Ra^{\gamma_u}$ and $Ra^{\gamma_{\theta^{\prime}}}$, respectively. In this $Ra$ number regime the thermal fluctuations close to the side wall are an order of magnitude larger than at the cell center due to the plumes that travel along the wall. 
Figure \ref{scale_uT} shows that the corresponding velocity scaling exponents increase smoothly from $\gamma_{u}=0.44$ in the cell center to $\gamma_{u}=0.49$ near the side wall. Figure \ref{scale_uT} shows that the corresponding temperature scaling exponent decreases from 
$\gamma_{\theta^{\prime}}=-0.18$ in the cell center to $\gamma_{\theta^{\prime}}=-0.20$ near the side wall. 

Table \ref{table1} summarizes the data for the scaling exponents available in the literature and compares them with the present ones. There are some differences among the values reported. This is due in part to the spatial dependence of this quantity, shown in Figure~\ref{scale_uT}, but also to the use of different experimental techniques which measure somewhat different quantities. Overall, there is a general consistency among the data shown. The origin of the residual differences cannot be ascertained on the basis of the presently available knowledge and must await further research.

\section{Statistics with respect to the orientation of the large scale circulation}

For $\Gamma=1$, the flow in the cell is characterized by a large scale circulation (LSC) ~\cite{krishnamurti81,sano89,sun03,ahlers09} as 
sketched in Figure~\ref{fig:plumes_sketch}. Most of the plumes travel in the LSC plane close to the side wall. 
In experiments the LSC orientation can be detected with thermistors embedded in the side wall~\cite{brown05} which measure relatively higher and  lower temperatures in the region of upflow and downflow~\cite{ahlers09}. Here we want to determine how the local heat flux depends on the location of measurement with respect to the LSC orientation plane. 

\subsection{Determination of LSC orientation}

We know from Ref.~\cite{cubaschumacher10} that, in the present range of $Ra$ numbers, the period of the LSC is approximately 10 dimensionless time units. On this basis, according to the last column of Table 1, our simulations are long enough to cover more than 400 LSC periods for the smallest $Ra$ and 175 for the highest one.

To determine the LSC orientation we use cosine fits of vertical velocity, temperature and heat flux over horizontal planes in the angular direction near the side wall at $r/L=0.45$. For this purpose we use the information from the numerical probes placed uniformly in this direction (see section 2)~\cite{ahlers09,stevens11}. 
In performing the fits we need to take into account that the orientation of the LSC plane changes with time. To this end we have pre-processed the instantaneous data by filtering them by means of short-time moving averages of four different durations, namely 4, 10, 20 and 50 dimensionless time units. 
We have fitted the results of each one of these time averages by an expression of the form  
\begin{equation}
f_i=f_m + A_f \cos(m\phi_i + \phi_{LSC}),
\label{eqcosine}
\end{equation}
where $f_i$ is the (short-time moving averaged) information provided by the $i^{th}$ probe at the angular position $\phi_i$,
$f_m$ is the angular mean value,  
$A_f$ is the amplitude of the fit and $\phi_{LSC}$ is the phase shift with respect to the reference frame of the computation; the integer $m$ equals 1 or 2 depending on the angular periodicity of the quantity $f$ as will be clear below.

\begin{figure}
\vspace*{2ex}
\includegraphics[width=0.4\textwidth]{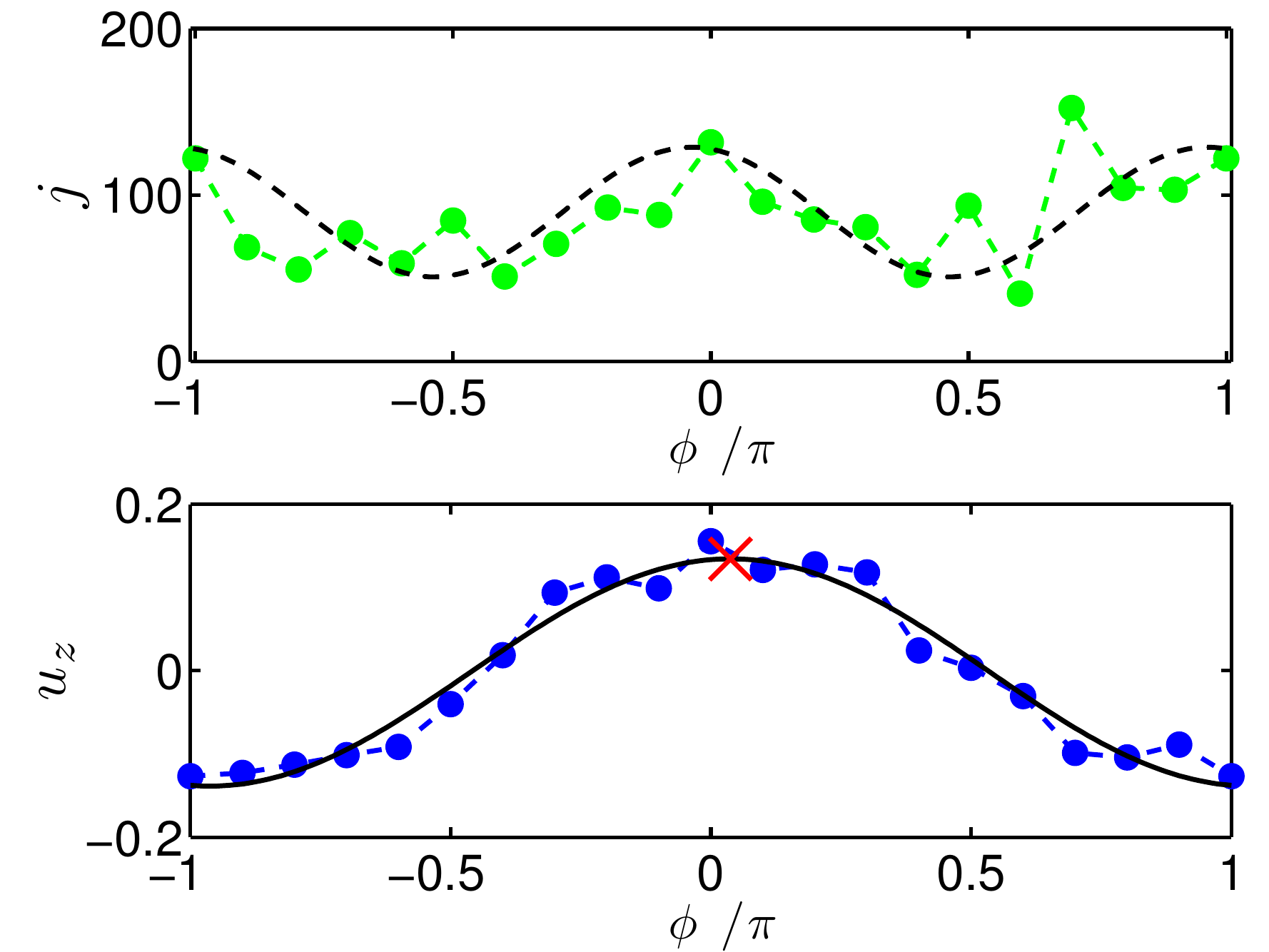} 
 \caption{Angular dependencies of heat flux and vertical velocity on the mid-plane at $r/L=0.45$ near the side wall for $Ra=2\times 10^8$: (a) instantaneous local heat flux (green points), (b) vertical velocity (blue points). The orientation of the LSC plane is determined by using the filtered time average of 20 dimensionless time units on vertical velocity data and the best cosine fit is shown in black solid line. 
The red cross indicates the maximum of the cosine fit, which we take as the LSC orientation.}
\label{fig:example_method}
\vspace*{-2ex}
\end{figure}

\begin{figure}
\vspace*{2ex}
\includegraphics[width=0.4\textwidth]{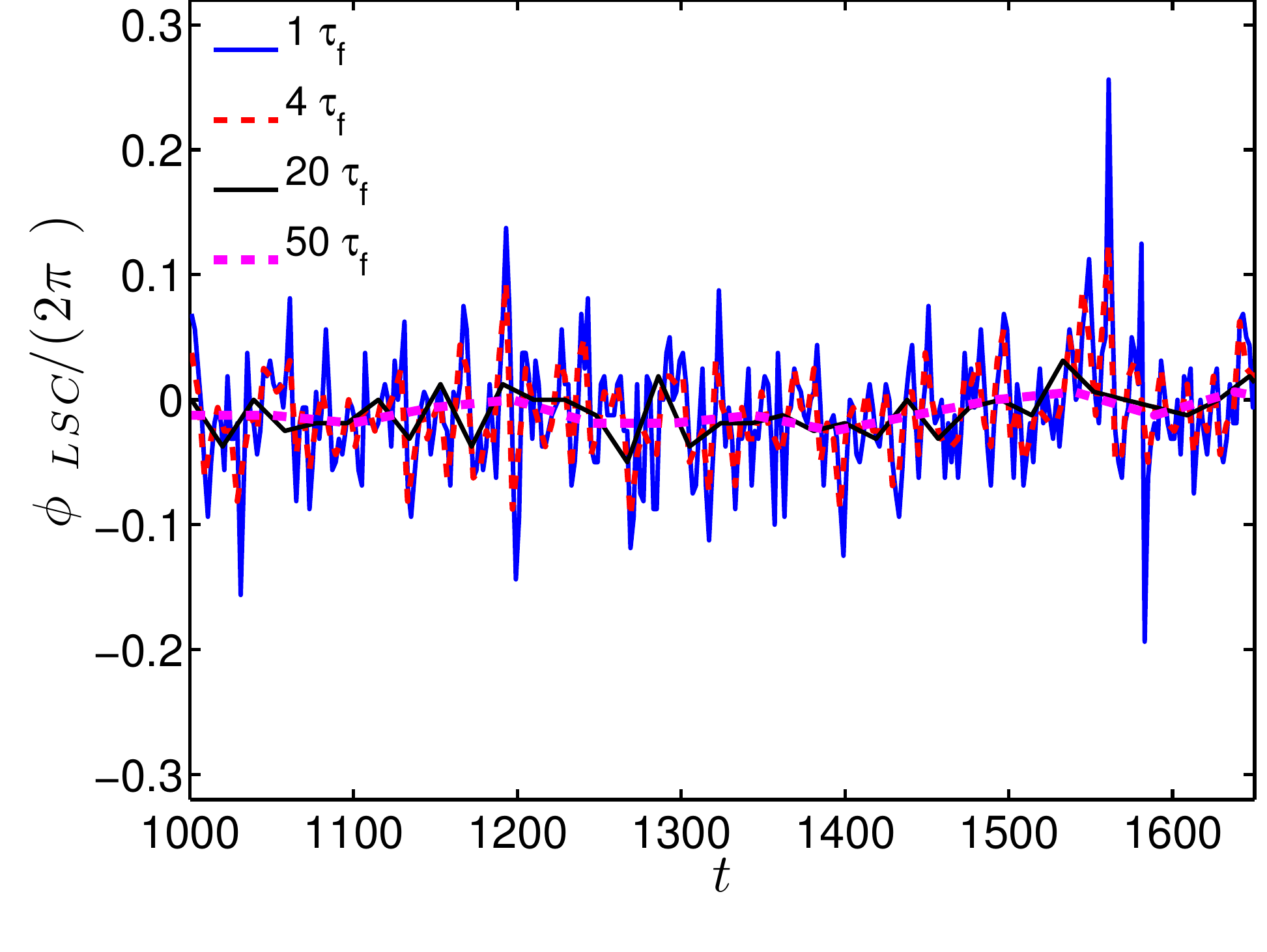} 
 \caption{LSC orientation as function of time for $Ra=2\times 10^8$. The time averaging that is applied to the signal before the analysis is 1 (blue-solid), 4 (red-dashdot), 20 (black-solid) and 50 (magenta-dash) dimensionless time units.
}
\label{fig:orientationvstime}
\vspace*{-2ex}
\end{figure}

An example of the difference between instantaneous and short-time-averaged data for the heat flux $j$ is shown in the top panel of Figure~\ref{fig:example_method} for $Ra=2\times 10^8$ on the mid-plane of the cylinder. The dashed line is the result of the fit to a portion of the instantaneous data averaged for 20 dimensionless time units. The bottom panel of Figure~\ref{fig:example_method} shows the filtered data and their cosine fit for the vertical velocity for the same positions and interval of time used for the top panel. The maximum of the curve is marked with a red cross and identifies the position of the LSC plane. A graph showing $\phi_{LSC}$ vs. time as calculated using the four different averaging times is shown in Figure~\ref{fig:orientationvstime}. As expected, the fluctuations in the position of the LSC plane are somewhat greater when short averaging times are used.

\begin{figure}
\vspace*{2ex}
\includegraphics[width=0.4\textwidth]{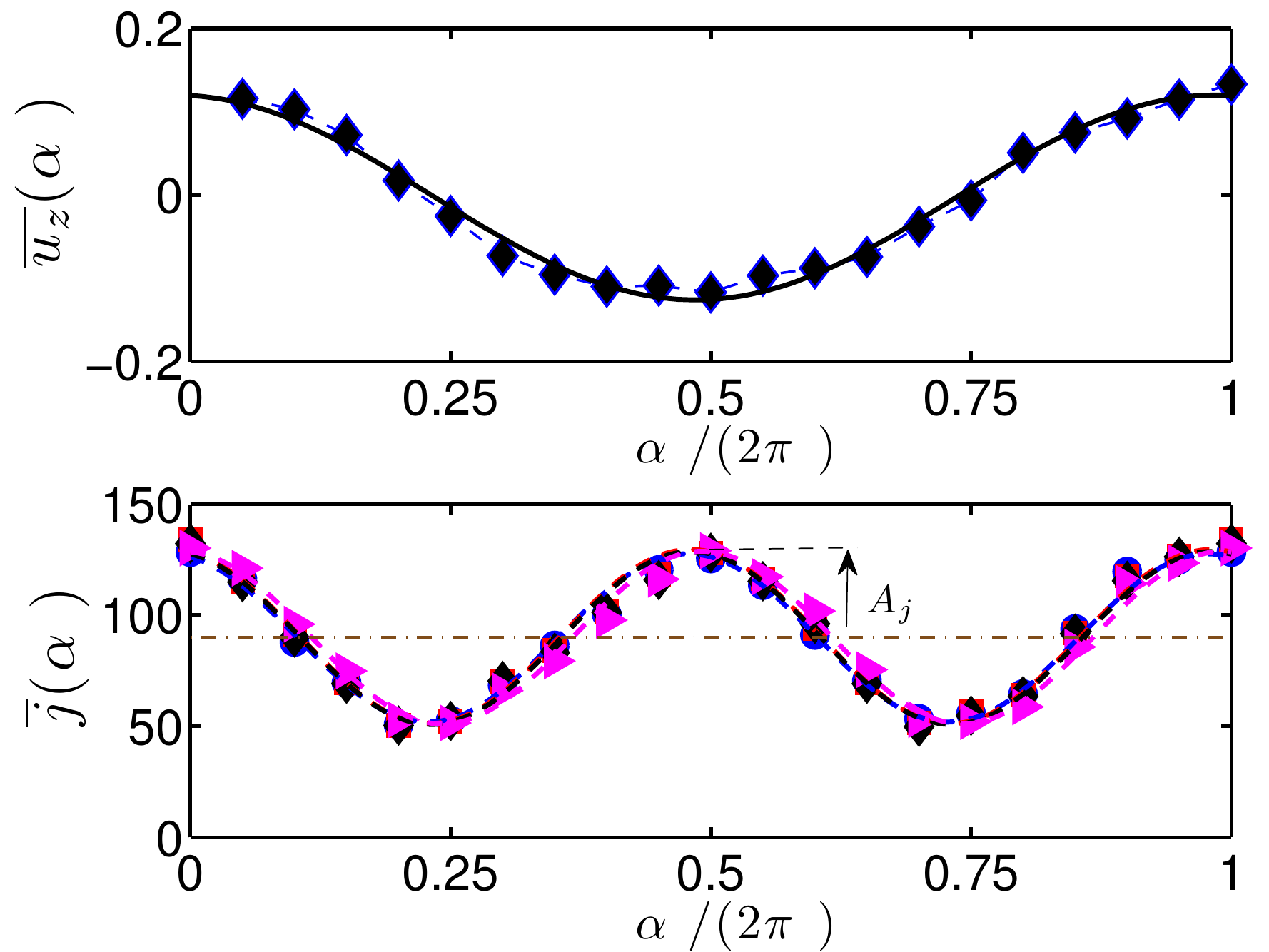} 
 \caption{Time averaged quantities relative to the orientation of the LSC plane near the side walls ($r/L$ = 0.45) at mid-height: (a) vertical velocity 
averaged over 20 dimensionless time units; (b) local heat flux $\overline{j}(\alpha)$ averaged over 4 (circle-blue), 10 (square-red), 20 (diamond-black) and 50 (triangle-magenta) dimensionless time units. The best cosine fit, with a period of 4$\pi$ which corresponds to $m=2$ in (\ref{eqcosine}), is shown by dashes. The mean heat flux (thin-dash dot) and the amplitude of variation $A_j$ are also shown. The hot plumes originate is in the neighborhood of 
$\alpha=0$.}
\label{fig:example_method2}
\vspace*{-2ex}
\end{figure}

Once the position of the LSC plane is known, all the short time averages can be referred to it by using a new angular co-ordinate $\alpha=\phi-\phi_{LSC}$. With this construction, the same value of $\alpha$ in each filtered segment of data identifies the same position relative to the  LSC and justifies taking the  average over all the filtered data sets. Such averages for the vertical velocity and heat flux at $r/L=0.45$ on the mid-plane are shown in Figure~\ref{fig:example_method2}.
A comparison of the two panels in this figure shows that the heat flux has double the periodicity of the velocity (i.e. the integer $m$ in equation (\ref{eqcosine}) equals 2 as opposed to $m=1$ for the velocity), because the heat flux is enhanced in correspondence of both the upward and downward moving streams of the LSC. The local heat flux is lower in the areas where the LSC is not present. 

The panel with the heat flux in Figure~\ref{fig:example_method2} actually contains four different sets of symbols, each one denoting the results obtained with a different averaging time. The virtual coincidence of these four sets demonstrates the robustness of the results with respect to the filtering time used.

\begin{figure}
\vspace*{2ex}
\includegraphics[width=0.4\textwidth]{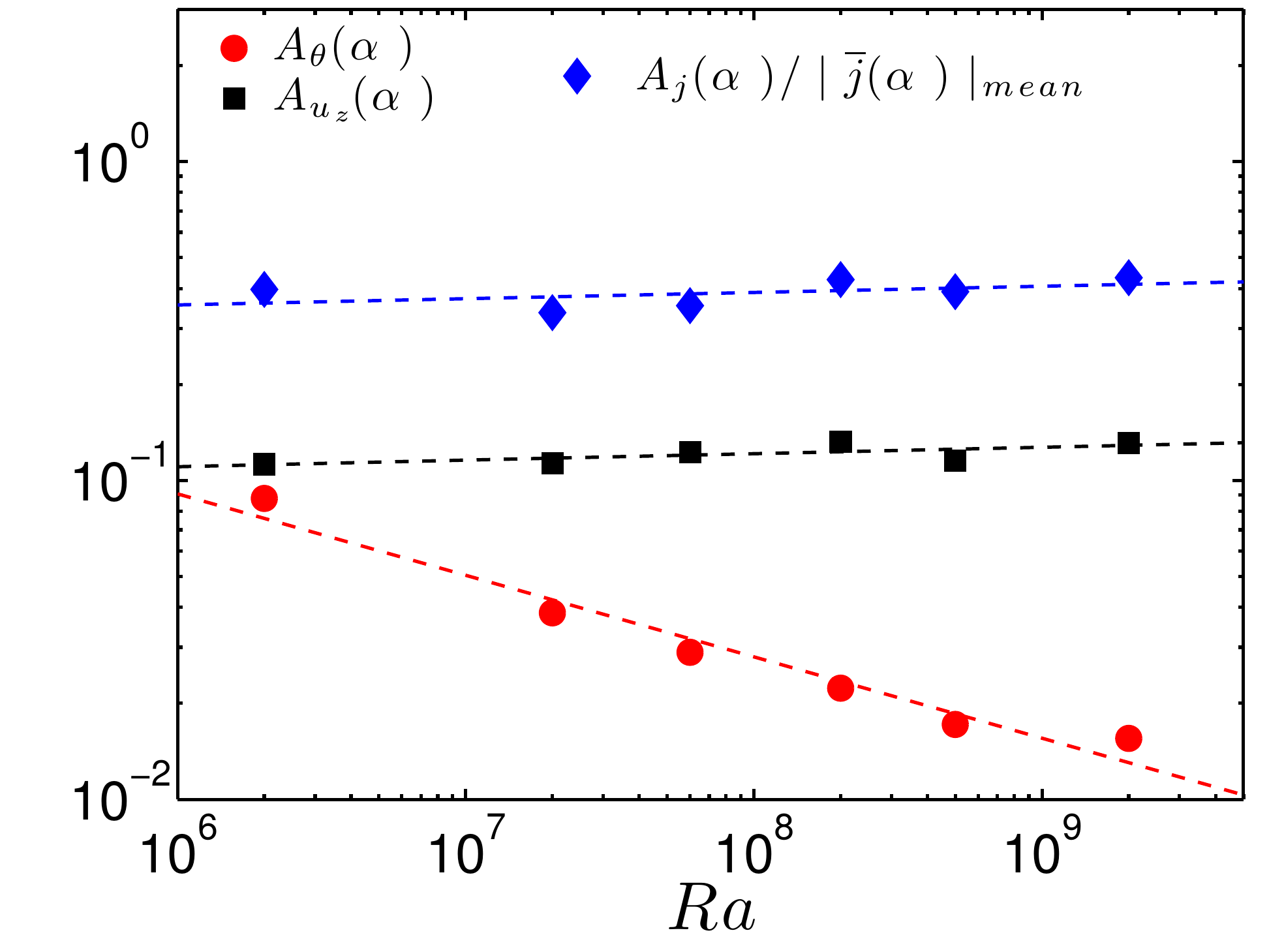} 
 \caption{The amplitude variation of cosine fits near the side walls, $r/L=0.45$, at the mid-height as function of $Ra$. The data are shown for temperature (circle-red), axial velocity (square-black) and normalized heat flux (diamond-blue). Here $\mid\overline{j}(\alpha)\mid_{mean}$ is the arithmetic mean heat flux in the azimuthal direction for a given $Ra$. The dashed lines indicates the power law fits for the data.}
\label{fig:amplitude}
\vspace*{-2ex}
\end{figure} 

In Figure~\ref{fig:amplitude} the amplitudes of the cosine fits for temperature, $A_{\theta}$, 
axial velocity, $A_{u_z}$, and normalized heat flux, $A_{j}$ are shown as functions of $Ra$ at mid-height near the side wall, $r/L=0.45$. Interestingly, these amplitudes have a power-law dependency on $Ra$, with scaling exponents $-0.250\pm0.010$, $0.020\pm0.005$, $0.019\pm0.005$ for temperature, axial velocity and local heat flux respectively.

\begin{figure}[!t]
\vspace*{-2ex}
    \subfigure{\includegraphics[width=0.4\textwidth]{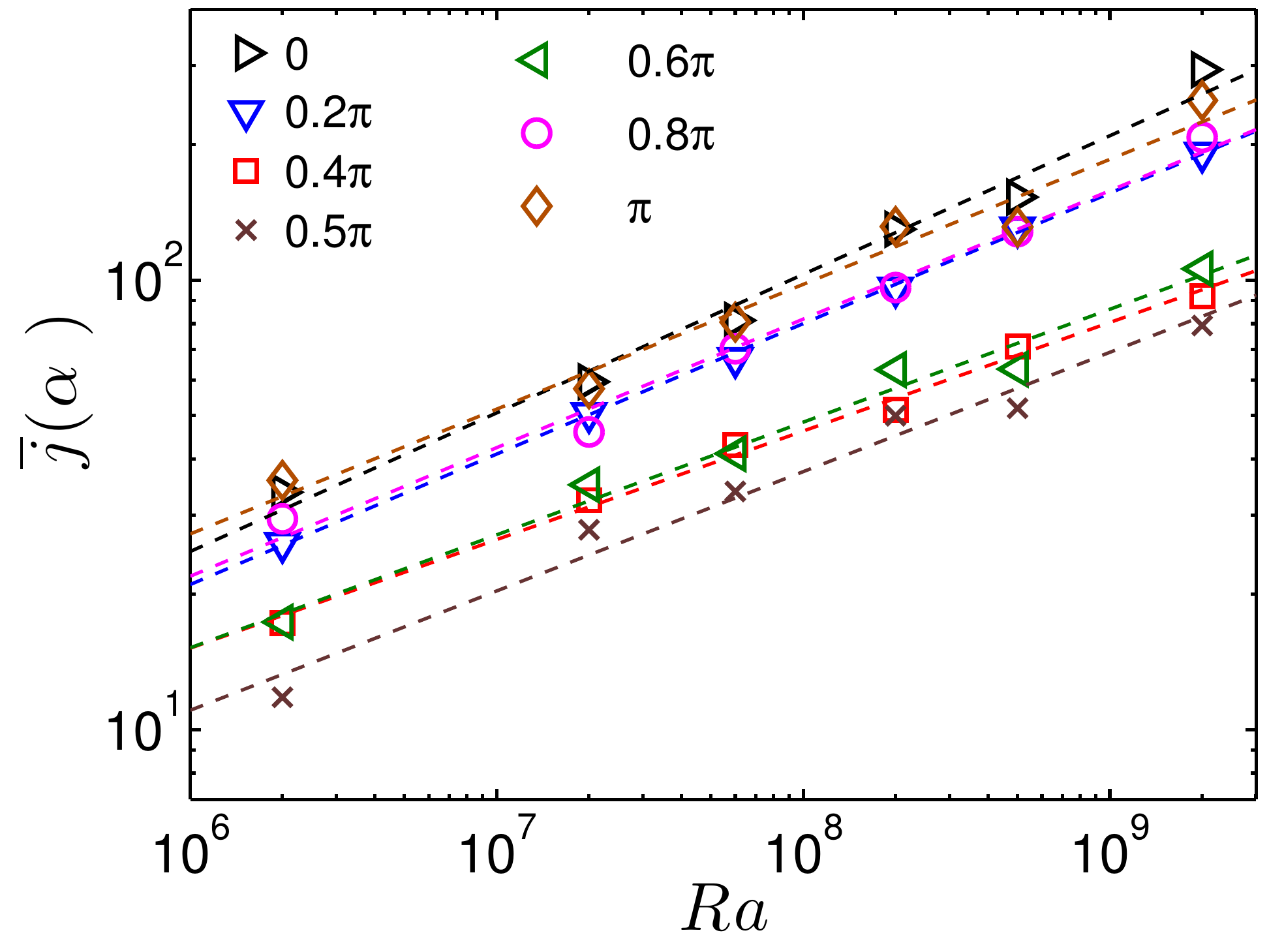}}
    \subfigure{\includegraphics[width=0.4\textwidth]{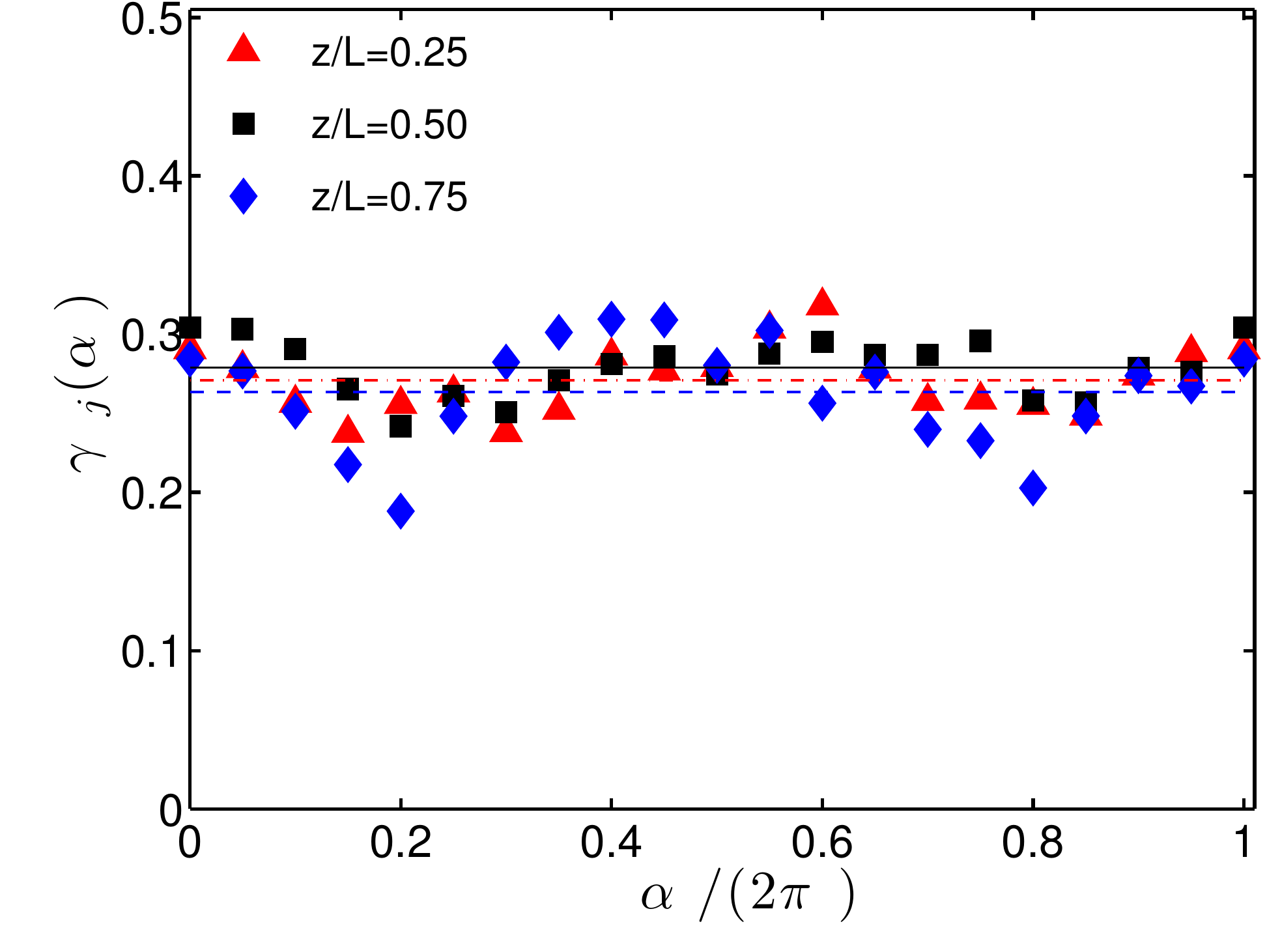}}
    \vspace*{-2ex}
  \caption{(a): Scaling of the time averaged local heat flux with $Ra$ near the side walls, $r/L=0.45$, at mid-height. 
The symbols indicate measurements taken at different $\alpha$. The dashed lines are the power law fits to the data. 
  (b) : Scaling exponent for the heat flux $\gamma_j(\alpha)$ relative to the LSC plane near the side wall, $r/L=0.45$ 
  at $z/L=0.25$ (triangles-red), $z/L=0.50$ (square-black), and $z/L=0.75$ (diamond-blue). 
  The straight lines indicate the arithmetic mean values of the scaling exponents.}
\label{fig:jz_powerlaw}
\vspace*{-2ex}
\end{figure}

Figure~\ref{fig:jz_powerlaw}a shows the time-averaged local heat flux as a function of $Ra$ at different $\alpha$ at mid-height near the side wall. From the figure it is clear that the local heat flux has a power law dependence on $Ra$. In agreement with the data in Figure~\ref{fig:amplitude} we find that the local heat flux at the LSC orientation increases faster than in the regions away from LSC. 
This is also revealed when the local heat flux scaling exponent as function of $\alpha$, see Figure~\ref{fig:jz_powerlaw}b, is considered.

\begin{figure}
\vspace*{2ex}
\includegraphics[width=0.4\textwidth]{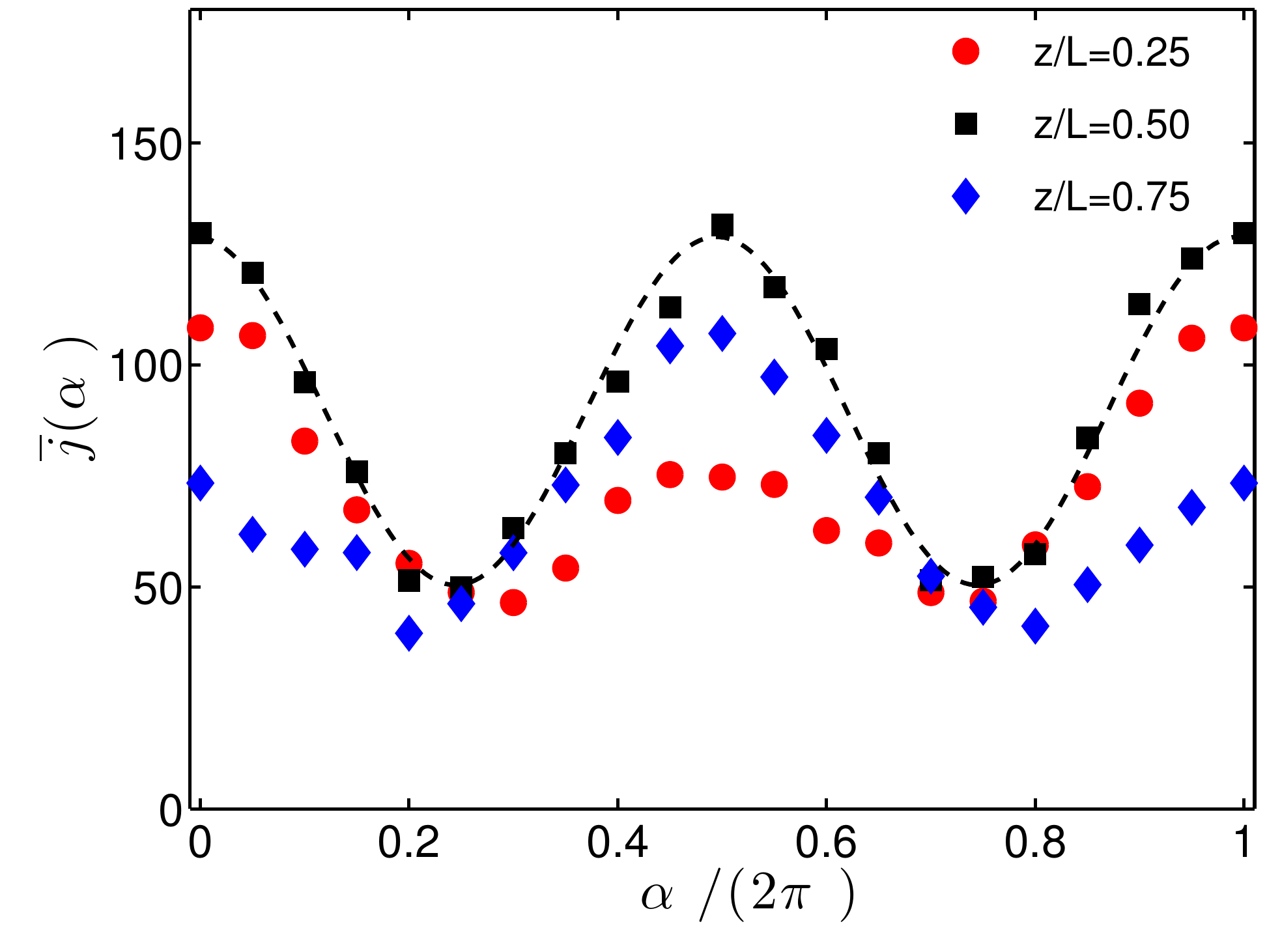}
 \caption{Time averaged local heat flux with respect to the LSC orientation for $Ra=2\times 10^8$ at different axial positions for $r/L=0.45$.  
 The points are, near the hot plate $z/L=0.25$ (circle-red), at mid-height $z/L=0.50$ (square-black), and near the top plate $z/L=0.75$ (diamond-blue). Moving averages are taken over 30 dimensionless time units.
}
\label{fig:jz_sections}
\vspace*{-2ex}
\end{figure}

Figure \ref{fig:jz_sections} shows the variation of the time-averaged local heat flux with $\alpha$ at three different heights. 
For $z/L=0.25$ larger values of $j$ occur near $\alpha=0$, and lower values near $\alpha=\pi$. 
A similar picture shifted by $\pi$ is found for $z/L=0.75$, with higher values near $\alpha=\pi$ and lower values near $\alpha=0$. 
At $z/L=0.50$, on the other hand, the levels at $\alpha=0$ and $\alpha=\pi$ are comparable.
These results suggest that the plane of the LSC is tilted as sketched in Figure~\ref{fig:plumes_sketch}. 
This feature has also been found in experiments of Funfschilling et al.~\cite{ahlers08}.

\section{Summary and Conclusions}

To summarize, we investigated numerically the scaling of the local heat flux in Rayleigh-B\'{e}nard convection of a fluid with $Pr=5.2$ (appropriate for water at 32 $^\circ$C) for $2\times10^6 \leq Ra \leq 2\times10^9$ in an unit aspect ratio cylinder. In this $Ra$ number regime the local heat flux is larger close to the side wall than on the axis. The local heat flux $u_z(T-T_0)$ is a positive quantity both when fluid warmer than the average temperature $T_0$ rises and fluid colder than $T_0$ sinks. The PDFs of the local heat flux have a positive skewness due to the dominance of plume transport in this $Ra$ range. 
On the mid-plane, the scaling exponents of the local heat flux with $Ra$ near the axis and close to the side wall agree well with the measurements of Shang et al.~\cite{shang08} and the predictions of Grossmann and Lohse \cite{gro04}. 
Here we have shown that these scaling exponents decrease monotonically with position $r$ from $0.43$ near the axis to $0.29$ close to the side wall. The scaling exponent for the Reynolds number based on the rms velocity depends on the radial position as well, with a value of $0.44$ near the axis and $0.49$ close to the side wall. For the scaling exponents of the temperature fluctuations we find $-0.18$ and $-0.20$ respectively.
%The scaling exponents for the temperature fluctuations is also $r$-dependent and decreases from $-0.18$ in the cell center to $-0.2$ close to the side wall. 
We showed the marked effect of the LSC which causes local heat fluxes more than twice as large in its plane than at $90^o$ from it.
This effect becomes stronger for high $Ra$.
%%%%%%%%%%%%%%%%%%%%%%%%%%%%%%%%%%%%%%%%%%
\section*{Acknowledgements}
We acknowledge the financial support by the Foundation for Fundamental Research on Matter (FOM) and the National Computing Facilities (NCF), both sponsored by NWO. This research is a part of industrial partnership program on Fundamentals of Heterogeneous Bubbly Flows (FHBF). 
The computations in this project have been performed on Huygens cluster of SARA in Amsterdam.
%%%%%%%%%%%%%%%%%%%%%%%%%%%%%%%%%%%%%%%%%%%%
%\bibliography{LocalHeatFlux_ref}% Produces the bibliography via BibTeX.

\begin{thebibliography}{10}
\bibitem{footnote}
The diffusive heat flux $-\kappa \overline{\partial_z T(r, t)}/(\kappa \Delta/L)$ is less than 1\% of the convective one, for all $r$ and $Ra$, due to the tiny temperature gradient at mid height.
\end{thebibliography}
%\bibliographystyle{prsty_withtitle}

\end{document}